\documentclass[journal,12pt,onecolumn,draftclsnofoot]{IEEEtran}
\IEEEoverridecommandlockouts

\pdfoutput=1

\usepackage[letterpaper, bottom=1in, right=1in, left=1in, top=1in]{geometry}
\usepackage{cite}
\usepackage{amsmath,amssymb,amsfonts,amsthm,bm}
\usepackage{graphicx}
\usepackage{textcomp}
\usepackage{xcolor}
\usepackage[font=footnotesize,skip=1pt]{caption}
\usepackage{subcaption}
\usepackage{enumitem}
\usepackage{cuted}
\usepackage{units}
\usepackage{multirow}
\usepackage[normalem]{ulem}
\usepackage[protrusion=true,expansion=true]{microtype}
\usepackage{balance}
\usepackage{mathtools}
\usepackage[linesnumbered,ruled,vlined]{algorithm2e}
\usepackage{algpseudocode}
\usepackage{pifont}

\graphicspath{{Pictures_Minor/}}

\theoremstyle{definition}
\newtheorem{theorem}{Theorem}

\newtheorem{example}{Example}

\allowdisplaybreaks
\DeclareMathOperator*{\argmin}{arg\,min}
\DeclareMathOperator*{\argmax}{arg\,max}
\def\BibTeX{{\rm B\kern-.05em{\sc i\kern-.025em b}\kern-.08em T\kern-.1667em\lower.7ex\hbox{E}\kern-.125emX}}

\begin{document}

\title{A Decentralized Pilot Assignment Algorithm for Scalable O-RAN Cell-Free Massive MIMO}

\author{
Myeung~Suk~Oh,~\IEEEmembership{Student~Member,~IEEE},
Anindya Bijoy Das,~\IEEEmembership{Member,~IEEE}, \\
Seyyedali~Hosseinalipour,~\IEEEmembership{Member,~IEEE}, Taejoon~Kim,~\IEEEmembership{Senior~Member,~IEEE}, \\
David~J.~Love,~\IEEEmembership{Fellow,~IEEE}, and Christopher~G.~Brinton,~\IEEEmembership{Senior~Member,~IEEE}
\thanks{M. S. Oh, A. B. Das, D. J. Love, and C. G. Brinton are with the
School of Electrical and Computer Engineering, Purdue University, West
Lafayette, IN, 47907 USA (e-mail: \{oh223, das207,
djlove, cgb\}@purdue.edu).}
\thanks{S. Hosseinalipour is with the Department of Electrical Engineering, University at Buffalo-SUNY, NY, 14260 USA (email: alipour@buffalo.edu).}
\thanks{T. Kim is with the Department of Electrical
Engineering and Computer Science, the University of Kansas, Lawrence, KS, 66045 USA
(email: taejoonkim@ku.edu).}
}

\maketitle

\begin{abstract}
    \vspace{-2mm}
    Radio access networks (RANs) in monolithic architectures have limited adaptability to supporting different network scenarios. Recently, open-RAN (O-RAN) techniques have begun adding enormous flexibility to RAN implementations. O-RAN is a natural architectural fit for cell-free massive multiple-input multiple-output (CFmMIMO) systems, where many geographically-distributed access points (APs) are employed to achieve ubiquitous coverage and enhanced user performance. In this paper, we address the decentralized pilot assignment (PA) problem for scalable O-RAN-based CFmMIMO systems. We propose a low-complexity PA scheme using a multi-agent deep reinforcement learning (MA-DRL) framework in which multiple learning agents perform distributed learning over the O-RAN communication architecture to suppress pilot contamination. Our approach does not require prior channel knowledge but instead relies on real-time interactions made with the environment during the learning procedure. In addition, we design a codebook search (CS) scheme that exploits the decentralization of our O-RAN CFmMIMO architecture, where different codebook sets can be utilized to further improve PA performance without any significant additional complexities. Numerical evaluations verify that our proposed scheme provides substantial computational scalability advantages and improvements in channel estimation performance compared to the state-of-the-art.
    \vspace{-3mm}
\begin{IEEEkeywords}
    \vspace{-2mm}
    Open-RAN (O-RAN), cell-free massive MIMO, deep reinforcement learning, pilot assignment. 
\end{IEEEkeywords}
\end{abstract}

\section{Introduction}\label{sec:introduction}

\subsection{Open Radio Access Network (O-RAN)}\label{ssec:intro_ORAN}

Next generation wireless technologies will likely employ many dispersed radio access networks (RANs) for ubiquitous coverage and enhanced user performance~\cite{Chowdhury20,Lee20}.
However, interconnecting different RANs to create one seamless network requires well-defined network functions and interfaces which are flexible in their integration capability.
Recently, the evolution of software-defined open RAN (O-RAN) solutions have added enormous flexibility to the implementation of current 5G networks~\cite{Singh20,Niknam22,Polese23} and development of emerging 6G networks.
O-RAN offers software-defined disaggregation on virtual network functions (VNFs) and necessary interfaces to support their coordination, allowing system implementations that are adaptive to various architectural settings.
With this openness and flexibility, O-RAN promotes interoperability across different RAN vendors and allows network operators to adapt to different wireless environments.

\begin{figure}[t]
    \centering
    \begin{subfigure}[!h]{0.49\textwidth}
    \centering
    \includegraphics[width=.9\linewidth]{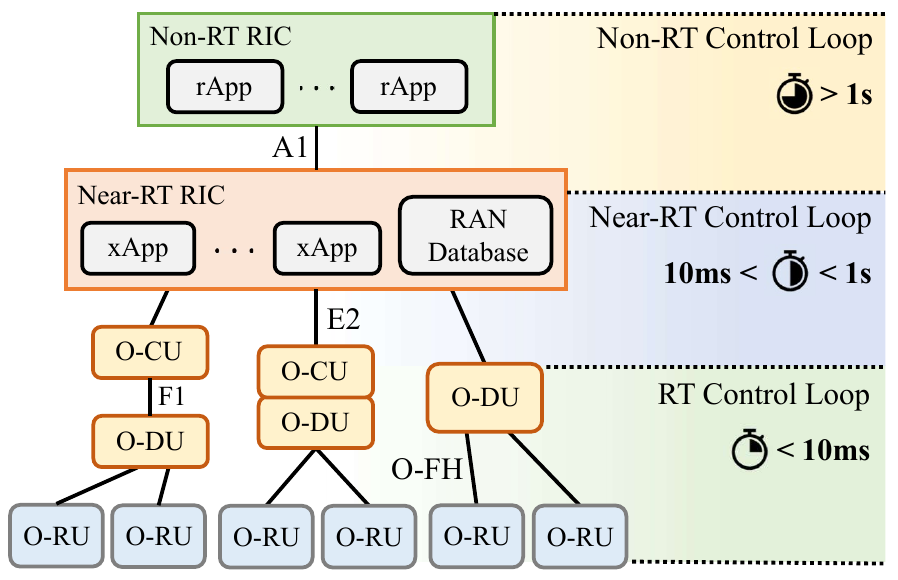}
    \caption{O-RAN architecture with different types of control loops.
    }
    \label{fig:ORAN_architecture}
    \end{subfigure}
    \begin{subfigure}[!h]{0.49\textwidth}
    \centering
    \includegraphics[width=0.78\linewidth]{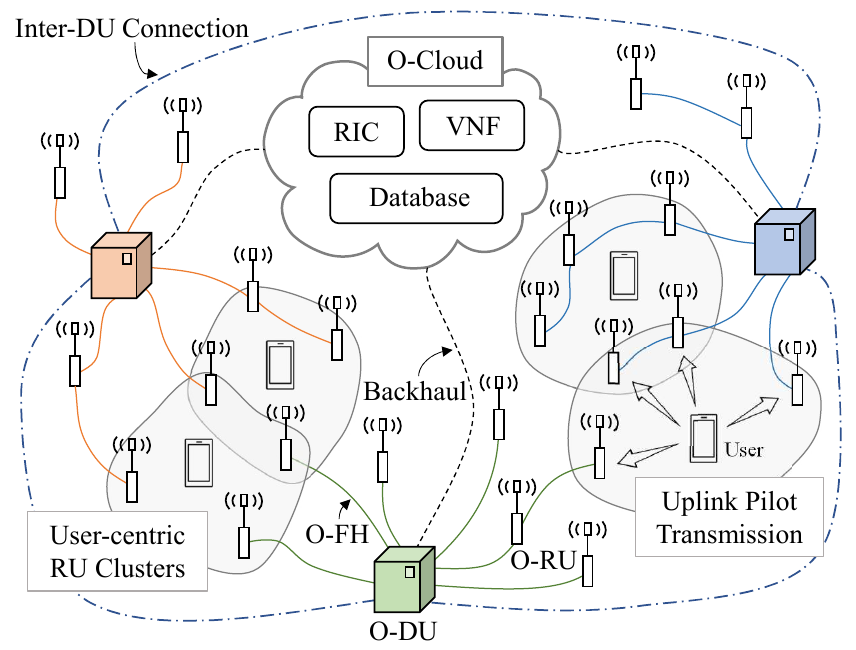}
    \caption{A decentralized CFmMIMO system realized in O-RAN.}
    \label{fig:system_layout}
    \end{subfigure}
    \caption{Illustrations of O-RAN architecture (left) and decentralized O-RAN CFmMIMO system (right).}
\end{figure}

O-RAN adopts the functional split defined in 3GPP~\cite{3GPP_NR} and defines three distinct units~\cite{ORAN}: the open central unit (O-CU), open distributed unit (O-DU), and open radio unit (O-RU). Moreover, O-RAN operation is divided into three different control loops~\cite{ORAN}: the real-time (RT), near-RT, and non-RT loops executing at different time-scales. The resulting O-RAN architecture and  standard names of interfaces between these elements, which enable practical implementation of many RAN operations, are depicted in Fig.~\ref{fig:ORAN_architecture}.

O-RAN offers two types of RAN intelligent controllers (RICs)~\cite{ORAN} as shown in Fig.~\ref{fig:ORAN_architecture}: near-RT RICs and non-RT RICs. Each of these RICs handles tasks manageable in different time-scales.
O-RAN offers virtualization of both RICs, which promotes flexibility in implementing data-driven intelligence tasks that will be key components of emerging wireless networks.
Various operations can be implemented via custom third-party applications called \emph{xApps}/\emph{rApps}~\cite{ORAN}, allowing RICs to be much more accessible to the public. In this work, we will consider the implementation of machine learning (ML) algorithms over these RICs to optimize pilot signal assignments.

Due to these aforementioned advantages offered by O-RAN, a number of opportunities to utilize O-RAN on future wireless technologies seem promising, some of which are:
\vspace{-1mm}
\begin{itemize}[leftmargin=4mm]
    \item \emph{Massive multiple-input multiple-output (MIMO) beamforming (BF)}: To implement ML-based BF strategies that handle both latency-sensitive (e.g., RT beam selection with quick updates) and data-intensive (e.g., policy update using a large dataset) tasks is challenging, and O-RAN provides a platform for these approaches~\cite{Mohsin21,Hewavithana22,ORAN_mimo}. 
    ML tasks are implemented in RICs, and BF operation can be split over O-RU and O-DU (e.g., option 7.2x~\cite{3GPP_split}) to maximize efficiency.
    \item \emph{Unmanned aerial vehicle (UAV) networking}: UAVs are typically deployed in dynamic environments (e.g., emergency rescue and aerial surveillance~\cite{Dao21}), where the network infrastructure is required to be extremely flexible and adaptive.
    Flexibility and interoperability offered by O-RAN can be exploited to meet this architectural need~\cite{Pham22,ORAN_use}.
    \item \emph{Localization via channel charting}: Channel charting is a data-driven localization technique~\cite{Studer18} that maps a user to radio geometry using channel information.
    For the practical implementation of channel charting, O-RAN can offer a balanced distribution of heavy computational load coming from the data that is consistently collected and updated for each user. 
\end{itemize}

\subsection{Cell-free Massive MIMO}\label{ssec:intro_cellfree}

One innovative idea to address the shortcomings of 5G cellular networks is to remove cell boundaries using many dispersed transmission/reception points.
This idea falls within the academic definition of cell-free massive MIMO (CFmMIMO)~\cite{Interdonato19_1,Zhang19,Zhang20}.
By deploying many geo-distributed access points (APs), CFmMIMO system alleviates the existing cell-edge problems by substantially improving both the reliability~\cite{Bjornson20_1} and energy efficiency~\cite{Yang18} compared to cellular massive MIMO.
These enhancements are due to the user-centric paradigm offered by CFmMIMO, where a group of APs are dynamically selected to form a cluster to serve each user.

In the early CFmMIMO literature, a system with APs connected to a single processing unit (PU) was considered for centralized operation.
However, in a scalable system where the number of users and APs grow large, the resulting complexity becomes prohibitive~\cite{Bjornson20_2}.
Thus, CFmMIMO with multiple decentralized PUs (Fig.~\ref{fig:system_layout}), each of which is connected to a disjoint subset of APs, has been introduced to consider scalability~\cite{Bjornson20_2,Interdonato19_2,He21,Ammar22_1}.
The decentralization allows the system to scale while still being practical by reducing the computational and fronthaul load on each PU~\cite{Zhang20}.
Nevertheless, implementing centralized CFmMIMO techniques (e.g., signal adaptation and resource allocation) into a decentralized architecture is a challenging task.

\subsection{CFmMIMO Pilot Assignment Problem}\label{ssec:intro_pilot}

In CFmMIMO, reliable channel estimation at both transmitter and receiver is absolutely critical to facilitate advanced diversity and signal processing techniques.
For channel estimation, a set of orthogonal pilots are used.
However, when the number of users grows beyond the number of available pilots, some users must share their pilots with others, leading to pilot contamination (PC) that can significantly degrade the channel estimation performance~\cite{Yin14}.
To cope with PC, various pilot assignment (PA) methods have been studied in the CFmMIMO literature~\cite{Ngo17,Sabbagh18,Chen21_1,Attarifar18,Liu20_2,Liu20_1,Buzzi21}.

In~\cite{Ngo17}, a greedy PA scheme with iterative pilot updates was proposed to mitigate PC.
A dynamic pilot reuse scheme to acquire a set of user-pairs for pilot sharing was proposed in~\cite{Sabbagh18}.
In~\cite{Chen21_1}, a user-group PA strategy, in which the same pilot is assigned to users with minimum overlapping APs, was proposed.
Other methods to solve the PA problem include k-means clustering~\cite{Attarifar18}, graph coloring~\cite{Liu20_2}, tabu-search~\cite{Liu20_1}, and Hungarian~\cite{Buzzi21} algorithms.

{These prior works~\cite{Ngo17,Attarifar18,Sabbagh18,Chen21_1,Liu20_1,Liu20_2,Buzzi21}, however, conduct PA via centralized processing. Thus, their computational complexities become prohibitive as the number of users grows large (e.g., Fig.~\ref{fig:complexity_user}).
While one can always consider conducting a set of uncoordinated local PAs, such an architecture without global orchestration can degrade the overall performance.
Successful decentralization requires a carefully designed coordination strategy to achieve performance comparable to the centralized case.
To the best of our knowledge, no work has yet developed and analyzed such a well-engineered distributed PA for CFmMIMO systems.}
In addition, these works~\cite{Ngo17,Attarifar18,Sabbagh18,Chen21_1,Liu20_1,Liu20_2,Buzzi21} use closed-form expressions derived from Bayesian estimation, requiring any relevant information (e.g., pathloss) to be known a priori.
The required information is in general obtained via estimation (e.g., pathloss can be estimated after collecting power measurements); however, for large-scale systems, especially under a dynamic environment, accurately estimated prior information is often not available due to the large overhead imposed, underscoring the need to develop a PA scheme that does not require prior knowledge.
{As a viable approach to address these issues, in this work, we adopt a learning-based optimization technique called multi-agent deep
reinforcement learning (MA-DRL) to conduct decentralized PA in a CFmMIMO system.}

\subsection{Overview of Methodology and Contributions}\label{ssec:intro_overview}

Motivated by the aforementioned challenges, we focus on PA in scalable CFmMIMO systems.
As CFmMIMO deploys a large number of APs for ubiquitous coverage, it is crucial to maintain a great level of implementation flexibility and interoperability across different RANs for scalability.
Hence, we propose to design our CFmMIMO system in O-RAN architecture.
As O-RAN  balances  operational complexities and computational loads via  a functional split along the network (i.e., O-RU/DUs and RICs), O-RAN becomes a natural solution for scalable CFmMIMO systems.

Based on the O-RAN CFmMIMO system, we formulate a decentralized PA problem and develop a learning-based PA scheme to solve it.
In doing so, we resort to a MA-DRL framework, in which a group of agents individually perform their learning to provide a low-complexity solution without an explicit training stage~\cite{Li21,Fredj22,Zhao21}.
Our PA scheme is designed to operate in the near-RT RIC of O-RAN.
We summarize the key contributions of our work below.
\begin{itemize}
    \item We design our CFmMIMO system based on the O-RAN architecture (Sec.~\ref{sec:system_model}). We specifically focus on channel estimation and pilot allocation models considering practical aspects (e.g., fronthaul overhead and operational complexity by each functional unit), which can be adopted to the O-RAN CFmMIMO systems.
    \item We design a Markov game model (Sec.~\ref{ssec:methodology_model}) for our MA-DRL which leads to an efficient solution for our decentralized PA problem.
    In particular, we formulate our reward based on observations that are directly measurable at the O-RUs.
    Thus, our scheme does not require prior knowledge of channel statistics, which is different from previous PA algorithms~\cite{Ngo17,Attarifar18,Sabbagh18,Chen21_1,Liu20_1,Liu20_2,Buzzi21}.
    \item Leverage the availability of RICs, we propose a novel learning-based PA scheme (Sec.~\ref{ssec:methodology_framework}) aiming to minimize the total mean squared error (MSE) across the users.
    By adopting the learning framework of MA-DRL, our scheme provides a low-complexity PA solution, the computation complexity of which increases at a much lower rate compared to the previous PA algorithms and therefore offers a scalability advantage to support large-scale systems.\vspace{-0.5mm}
    \item Utilizing the decentralization of our system, we consider two effective ways to improve the PA performance: (i) inter-DU message passing for observation sharing and (ii) low-complexity codebook search (CS) algorithm (Sec.~\ref{ssec:methodology_codebook}) that jointly operates with our PA scheme.
    Numerical results verify that these approaches can further improve the PA performance.
    \item We show that our PA scheme can maintain its performance over a mobile environment, which is possible due to (i) the DRL framework that naturally performs adaptive learning and (ii) the CS algorithm with iterative greedy search.
    Previous PA methods only consider a static environment and do not address the user mobility. 
    \item We numerically evaluate (Sec.~\ref{sec:simulation}) the performance of our PA scheme against the state-of-the-art~\cite{Liu20_1,Buzzi21} in both channel estimation performance and computational complexity.
    The results show that our scheme outperforms the benchmarks in terms of sum-MSE and scalability.
\end{itemize}

\section{System Model and Problem Formulation}\label{sec:system_model}

In this section, we first describe the architecture of the considered O-RAN-based CFmMIMO system (Sec.~\ref{ssec:system_cellfree}) to establish a foundation for CFmMIMO decentralization.
Then, after describing the channel model (Sec.~\ref{ssec:system_channel}), we provide details of codebook-based channel estimation (Sec.~\ref{ssec:system_estimation}) and uplink/downlink data transmission (Sec.~\ref{ssec:system_transmission}).
Finally, we formulate our decentralized PA problem (Sec.~\ref{ssec:system_formulation}) and explain the relationship between the PA task and channel estimation performance.

\subsection{CFmMIMO Configuration in O-RAN Architecture}\label{ssec:system_cellfree}

Our decentralized O-RAN CFmMIMO system is illustrated in Fig.~\ref{fig:system_layout}.
We consider $M$ single-antenna O-RUs and $U$ O-DUs collected in sets $\mathcal{M} = \{1,2, \ldots, M\}$ and $\mathcal{U} = \{1,2, \ldots, U\}$, respectively.
The O-RUs are randomly placed using uniform distribution across the area that is divided into $U$ disjoint regions for system decentralization, and each O-DU is deployed to one of the regions.
Each O-RU is connected to one of the O-DUs in $\mathcal{U}$ via an open fronthaul (O-FH) connection such that O-RUs within each subdivided region are connected to the same O-DU.
We define $\mathcal{M}^\mathsf{DU}_{u} \subseteq \mathcal{M}$ as the set of O-RUs connected to O-DU $u \in \mathcal{U}$.
We assume inter-DU connections~\cite{Ranjbar22} to form RU clusters that are fully user-centric since the users can be served by RUs from different sets of $\mathcal{M}^\mathsf{DU}_{u}$.
We assume ideal O-FH and inter-DU connections~\cite{Liu20_1,Liu20_2}.

Here, we have our O-DUs connected to O-Cloud~\cite{ORAN} via backhaul network (Fig.~\ref{fig:system_layout}).
O-Cloud is the cloud computing platform that supports the virtualized network functions (VNFs) within O-RAN, which include RICs. 
In designing our PA scheme, we specifically focus on the near-RT RIC that communicates with O-DUs via E2 interface (Fig.~\ref{fig:ORAN_architecture}).
Within the near-RT RIC, we assume $U$ independent learning agents, each of which has a one-to-one correspondence to one of the O-DUs in the system.
Note that we assume multiple agents to fully impose decentralization on our system.
Each agent in the near-RT RIC conducts local learning through the O-DU and O-RUs connected.
We also consider a single non-RT RIC interacting with the near-RT RIC via an A1 interface (Fig.~\ref{fig:ORAN_architecture}), which is responsible for learning model updates of the near-RT RIC.

Next, we consider $K$ single-antenna users in a set $\mathcal{K}=\{1, 2, \ldots, K\}$.
For each user $k$, a user-centric RU cluster is formed such that only $M^\mathsf{UE}_k \ll M$ O-RUs are engaged to serve the user, where we define $\mathcal{M}^\mathsf{UE}_{k} \subset \mathcal{M}$ to be the set of O-RUs serving user $k \in \mathcal{K}$ (i.e., $M^\mathsf{UE}_{k} = |\mathcal{M}^\mathsf{UE}_{k}|$ where $|\cdot|$ denotes the set cardinality).
Each $\mathcal{M}^\mathsf{UE}_{k}$ is assumed to be selected and updated using a procedure independent from our PA 
(e.g., radio resource control (RRC) setup procedure~\cite{3GPP_RRC}).
We also define $\mathcal{K}^\mathsf{RU}_m \subset \mathcal{K}$ to be the set of users served by O-RU $m \in \mathcal{M}$.

Since we have $U$ multiple agents performing PA, each user $k\in\mathcal{K}$ must belong to one of these agents.
To develop user-to-agent pairings, we consider two different types of users:
(i) user $k$ whose $\mathcal{M}^\mathsf{UE}_{k}$ is connected to a single O-DU $u$, i.e., $\mathcal{M}^\mathsf{UE}_{k} \subseteq \mathcal{M}^\mathsf{DU}_{u}$, which we simply pair that user $k$ to the corresponding agent $u$, and (ii) user $k$ whose $\mathcal{M}^\mathsf{UE}_{k}$ consists of O-RUs from different O-DUs.
For the second type, a serving O-DU~\cite{Ranjbar22}, which can be defined by any reasonable criterion (e.g., the O-DU with the most number of O-RUs serving the user), is determined and paired with the user.
We define $\mathcal{K}^\mathsf{DU}_u$ to be the set of users whose PA is managed by O-DU $u$.

\begin{figure}[!t]
    \centering
    \includegraphics[width=.6\linewidth]{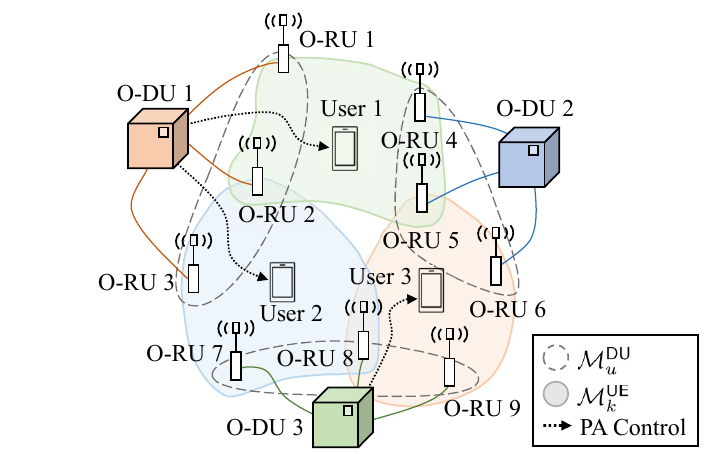}
    \caption{A list of our defined sets and their visual examples for the given decentralized cell-free O-RAN layout.}
    \label{fig:set_visuals}
\end{figure}

\begin{example}
We consider a scenario with $U=3$, $M=9$, and $K=3$, which is illustrated in Fig.~\ref{fig:set_visuals}.
For decentralization, each O-DU is connected to three O-RUs that are closest (e.g., $\mathcal{M}^\mathsf{DU}_1=\{1,2,3\}$), and user-centric RU clusters with $M^\mathsf{UE}_k=4$ are formed for each user (e.g., $\mathcal{M}^\mathsf{UE}_1=\{1,2,4,5\}$).
Note that an O-RU can serve multiple users (e.g., $\mathcal{K}^\mathsf{RU}_2=\{1,2\}$).
Since each user needs an agent for PA, the user is paired to one of the three O-DUs (e.g., $\mathcal{K}^\mathsf{DU}_1=\{1,2\}$).
\end{example}

\subsection{Time-varying Channel Model}
\label{ssec:system_channel}

We assume a periodic channel estimation with time interval $T_e$ and indicate each estimation instance using index $i=0,1, \ldots,N$.
The channel between user $k \in \mathcal{K}$ and O-RU $m \in \mathcal{M}$ during channel estimation instance $i$ is formally expressed as
\begin{equation}
    g_{km}^{(i)}=\sqrt{\beta_{km}^{(i)}}h^{(i)}_{km},
    \label{eq:channel}
\end{equation}
where $h_{km}^{(i)} = \mu_k h_{km}^{(i-1)} + \sqrt{(1-\mu_k^{2})}n^{(i)}_{km}$ is the small-scale fading factor following a first-order time-varying Gauss-Markov process for $i=1,2,\ldots,N$.
The perturbation terms $\{n^{(i)}_{km}\}$ are  zero-mean, unit-variance complex Gaussian random variables that are  independent and identically distributed (i.i.d.) over $k$, $m$, and $i$, i.e., $n_{km}^{(i)} \sim \mathcal{CN}(0,1)$.
At $i=0$, we assume $h_{km}^{(0)} \sim \mathcal{CN}(0,1)$ to be mutually independent from $n_{km}^{(1)}$.
The correlation coefficient $\mu_k$ for user $k$ is defined as $\mu_k=J_0(2\pi\frac{v_k}{c}f_cT_e)$~\cite{Kim11}, where $J_0(\cdot)$ is the Bessel function of the first kind of order zero, $v_k$ is the velocity of user $k$, $f_c$ is the carrier frequency, and $c=3\times10^8$ m/s is the speed of light.
The magnitude of $h^{(i)}_{km}$ is designed to follow a Rayleigh distribution, which is effective in modeling a dense scattering wireless environment~\cite{MIMO}.
The term $\beta_{km}^{(i)}$ in~\eqref{eq:channel} is the large-scale fading factor that is inversely proportional to the distance between user $k$ and O-RU $m$ at the channel estimation instance $i$.
There exist multiple realistic large-scale fading models, including the 3GPP urban-micro line-of-sight pathloss model~\cite{3GPP} that is used in our numerical evaluations.

\subsection{Codebook-based Channel Estimation} \label{ssec:system_estimation}

We consider uplink channel estimation with $T_p$ dedicated channel uses for each estimation instance, allowing $T_p$ orthogonal pilots to be available.
For each instance $i$, user $k \in \mathcal{K}^\mathsf{DU}_u$ is assigned with one of the $T_p$ pilots in a mutually orthogonal codebook $\mathcal{T}_u^{(i)} = \{\boldsymbol{\phi}_{u,1}^{(i)},\boldsymbol{\phi}_{u,2}^{(i)},\ldots,\boldsymbol{\phi}_{u,T_p}^{(i)}\}$, where each $\boldsymbol{\phi}_{u,t}^{(i)}$ for $t = 1, 2, \dots, T_p$ is a unit-norm complex vector of length $T_p$.
Thus, for $t,t'=1,2,\ldots,T_p$,  $(\boldsymbol{\phi}_{u,t}^{(i)})^{\text{H}}\boldsymbol{\phi}^{(i)}_{u,t'} = 1$ if $t = t'$, and {\it zero} otherwise, where $(\cdot)^\text{H}$ denotes the conjugate transpose.
We denote the pilot assigned to user $k$ as $\boldsymbol{x}^{(i)}_{k}$.

To conduct channel estimation, each user $k \in \mathcal{K}$ transmits the assigned pilot $\boldsymbol{x}^{(i)}_{k}$.
The signal vector (of length $T_p$) received by O-RU $m \in \mathcal{M}$ is then expressed as
$\boldsymbol{y}^{(i)}_m = \mathbf{X}^{(i)}\boldsymbol{g}^{(i)}_m + \boldsymbol{w}^{(i)}_m = \sum_{k\in\mathcal{K}}g^{(i)}_{km}\boldsymbol{x}^{(i)}_{k}+\boldsymbol{w}^{(i)}_m$, where $\mathbf{X}^{(i)}\hspace{-1mm}=[\boldsymbol{x}^{(i)}_{1} \boldsymbol{x}^{(i)}_{2} \cdots \boldsymbol{x}^{(i)}_{K}]$ is the $T_p \times K$ pilot matrix and $\boldsymbol{g}^{(i)}_{m}\hspace{-1mm}=[g^{(i)}_{1m} \; g^{(i)}_{2m} \cdots g^{(i)}_{Km}]^\top$ is the channel vector (of length $K$) for O-RU $m$. Here, $\boldsymbol{w}^{(i)}_m \sim \mathcal{CN}(\mathbf{0},\sigma^2\mathbf{I}_{T_p})$ is the zero-mean complex Gaussian noise vector of length $T_p$ with covariance $\sigma^2\mathbf{I}_{T_p}$, where $\mathbf{I}_n$ is the $n \times n$ identity matrix.

\begin{figure}[t]
    \centering
    \begin{subfigure}[!h]{0.49\textwidth}
    \centering
    \includegraphics[width=1\linewidth]{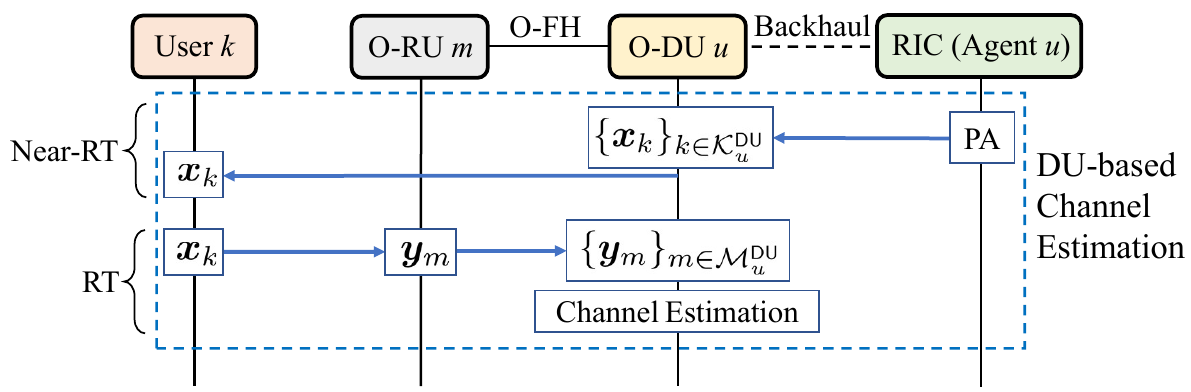}
    \caption{DU-based channel estimation.}
    \label{fig:CE_structure_DU}
    \end{subfigure}
    \begin{subfigure}[!h]{0.49\textwidth}
    \centering
    \includegraphics[width=1\linewidth]{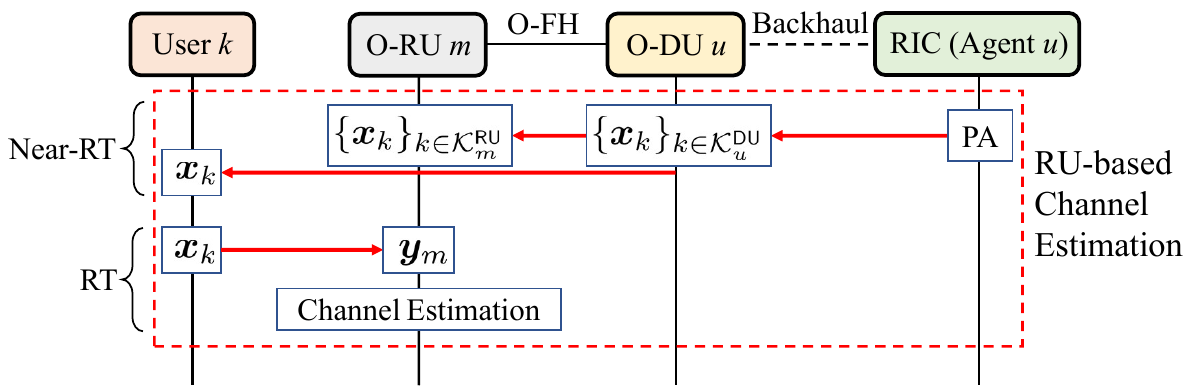}
    \caption{RU-based channel estimation.}
    \label{fig:CE_structure_RU}
    \end{subfigure}
    \caption{A block diagram of two different channel estimation structures.}
    \label{fig:CE_structure}
\end{figure}

We discuss two different channel estimation structures within O-RAN architecture, which we illustrate in Fig.~\ref{fig:CE_structure}, and compare their communication overhead by computing the number of bits exchanged during a single near-RT control loop.
One structure (Fig.~\ref{fig:CE_structure_DU}) performs channel estimation at O-DU whereas the estimation occurs at O-RU in the other structure (Fig.~\ref{fig:CE_structure_RU}).
We assume that $N_\mathsf{n}$ RT loops occur for each near-RT loop.

Suppose representing the PA information $\boldsymbol{x}^{(i)}_k$ and received signal $\boldsymbol{y}^{(i)}_m$ requires $b_\mathsf{d} = \log_2T_p$ and $b_\mathsf{u} = 2BT_p$ bits, respectively, where $2B$ is the number of bits used to represent a complex number.
For DU-based channel estimation, the RIC first sends out the PA information of the controlled users $\{\boldsymbol{x}^{(i)}_k\}_{k\in\mathcal{K}^\mathsf{DU}_u}$ over the backhaul using $b_\mathsf{d}\vert\mathcal{K}^{\mathsf{DU}}_u\vert$ bits.
O-DU $u$ then passes this pilot information to the master O-RU that serves user $k$ over the O-FH using another $b_\mathsf{d}\vert\mathcal{K}^{\mathsf{DU}}_u\vert$ bits.
For each of the $N_\mathsf{n}$ RT loops, $\boldsymbol{y}^{(i)}_m$ from each O-RU $m\in\mathcal{M}^\mathsf{DU}_u$ must be collected by the O-DU, which results in $b_\mathsf{u}\vert\mathcal{M}^{\mathsf{DU}}_u\vert N_\mathsf{n}$ bits exchanged over the O-FH.
Hence, the total amount of overhead for DU-based channel estimation is  $\sum_{u=1}^U\big(2b_\mathsf{d}\vert\mathcal{K}^{\mathsf{DU}}_u\vert + b_\mathsf{u}\vert\mathcal{M}^{\mathsf{DU}}_u\vert N_\mathsf{n}\big)$ bits.
Note that $2b_\mathsf{d}\vert\mathcal{K}^{\mathsf{DU}}_u\vert$ accounts for the data transferred in both backhaul and O-FH links.

For the RU-based channel estimation, each of the O-RUs serving user $k$ must be informed with the pilot information $\boldsymbol{x}^{(i)}_k$ to conduct channel estimation.
Therefore, in addition to the $b_\mathsf{d}\vert\mathcal{K}^{\mathsf{DU}}_u\vert$ bits exchanged between the RIC and O-DU over the backhaul, $\sum_{m\in\mathcal{M}^\mathsf{DU}_u}b_\mathsf{d}\vert\mathcal{K}^{\mathsf{RU}}_m\vert$ bits must be transferred via O-FH to deliver the pilot information to the O-RUs.
Note that, for RU-based channel estimation, no RT data transfer is required since the estimation occurs at each O-RU.
Hence, the total amount of overhead for RU-based channel estimation is given by $b_\mathsf{d}\sum_{u=1}^U\big(\vert\mathcal{K}^{\mathsf{DU}}_u\vert + \sum_{m\in\mathcal{M}^\mathsf{DU}_u}\vert\mathcal{K}^{\mathsf{RU}}_m\vert\big)$.
Table~\ref{tb:overhead} shows the amount of overhead in bits per near-RT loop required for channel estimations when $M=96$, $U=4$, $M^\mathsf{UE}_k=8$, and $B=8$.
From the result, we confirm that RU-based estimation imposes less overhead than DU-based one does.
Note that it also benefits from latency advantage as no data exchange is needed for RT loops.
Hence, similar to the work in~\cite{Ngo17}, we assume our channel estimation to take place at O-RUs.

Next, in case of user-centric RU clustering, each RU $m \in \mathcal{M}$ only needs to estimate $|\mathcal{K}^\mathsf{RU}_m|$ different channels (i.e., $\{g^{(i)}_{km}\}_{k\in\mathcal{K}^\mathsf{RU}_m}$) associated with users in $\mathcal{K}^\mathsf{RU}_m$.
For estimating the channel, we consider two different techniques called \emph{pilot-matching}~\cite{Bjornson20_1} and \emph{least-square}~\cite{Liu14} estimations.
If we set $\widehat{\boldsymbol{g}}^{(i)}_{m}=[\widehat{g}^{(i)}_{km}]^\top_{k \in \mathcal{K}^\mathsf{RU}_m}$ as the $|\mathcal{K}^\mathsf{RU}_m|$-length estimated channel vector from O-RU $m$ during the channel estimation instance $i$, pilot-matching and least-square estimations are expressed as
\begin{equation}
     \widehat{\boldsymbol{g}}^{(i)}_{m} = (\bar{\mathbf{X}}_{m}^{(i)})^\text{H}\boldsymbol{y}^{(i)}_m = (\mathbf{Z}_{m}^{(i)})^\text{H}(\mathbf{X}^{(i)})^\text{H}\boldsymbol{y}^{(i)}_m
     \label{eq:pilot_matching}
\end{equation}
and
\begin{equation}
    \widehat{\boldsymbol{g}}^{(i)}_{m} = (\bar{\mathbf{X}}_{m}^{(i)})^\text{H}(\mathbf{X}^{(i)}(\mathbf{X}^{(i)})^\text{H})^{-1}\boldsymbol{y}^{(i)}_m = (\mathbf{Z}_{m}^{(i)})^\text{H}(\mathbf{X}^{(i)})^\text{H}(\mathbf{X}^{(i)}(\mathbf{X}^{(i)})^\text{H})^{-1}\boldsymbol{y}^{(i)}_m,
    \label{eq:least_square}
\end{equation}
respectively, where $\bar{\mathbf{X}}_{m}^{(i)}=\mathbf{X}^{(i)}\mathbf{Z}_{m}^{(i)}=[\boldsymbol{x}^{(i)}_{k}]_{k \in \mathcal{K}^\mathsf{RU}_m}$ is the $T_p \times |\mathcal{K}^\mathsf{RU}_m|$ pilot matrix of the users served by O-RU $m$. We define a $K\times|\mathcal{K}^\mathsf{RU}_m|$ selection matrix $\mathbf{Z}_{m}^{(i)}=[\boldsymbol{z}^{(i)}_{k}]_{k \in \mathcal{K}^\mathsf{RU}_m}$ where $\boldsymbol{z}^{(i)}_{k}$ is the $K$-length unit-vector with its $k$-th element being \textit{one}.
Now, when some of $K$ users share the pilot, $\mathbf{X}^{(i)}$ is not unitary (i.e., $(\mathbf{X}^{(i)})^\text{H}\mathbf{X}^{(i)}\neq\mathbf{I}_{K}$), so the least-square estimation in~\eqref{eq:least_square}, which utilizes the pseudo-inverse term $(\mathbf{X}^{(i)})^\text{H}(\mathbf{X}^{(i)}(\mathbf{X}^{(i)})^\text{H})^{-1}$ to negate the PC, yields better estimation performance.
However, in the least-square approach, since $\mathbf{X}^{(i)}$ needs to be known to every O-RU and the size of $\mathbf{X}^{(i)}$ increases linearly with $K$, the resulting overhead causes significant delay as the number of users grows.
Note that, for the case of pilot-matching, each O-RU $m$ only needs to know $\{\boldsymbol{x}^{(i)}_k\}_{k\in\mathcal{K}^\mathsf{RU}_m}$ to obtain $\bar{\mathbf{X}}_{m}^{(i)}$.
This motivates the pilot-matching channel estimation scheme in~\eqref{eq:pilot_matching} for scalability~\cite{Bjornson20_1}.
The estimated channel $\widehat{g}^{(i)}_{km}$ is then expressed as
\begin{align}
    \hspace{-2mm}\widehat{g}^{(i)}_{km}\hspace{-1mm}=\hspace{-0.5mm}(\hspace{-0.5mm}\boldsymbol{x}_{k}^{(i)})^\text{H}\boldsymbol{y}^{(i)}_m \hspace{-0.5mm}=\hspace{-1.5mm}\sum_{k'\in\mathcal{K}}g^{(i)}_{k'm}(\hspace{-0.5mm}\boldsymbol{x}_{k}^{(i)})^\text{H}\boldsymbol{x}^{(i)}_{k'}\hspace{-0.5mm}+\hspace{-0.5mm}(\hspace{-0.5mm}\boldsymbol{x}_{k}^{(i)})^\text{H}\boldsymbol{w}^{(i)}_m \hspace{-0.5mm}=\hspace{-0.5mm}g^{(i)}_{km}\hspace{-0.5mm}+\hspace{-1.5mm}\sum_{\substack{k'\in\mathcal{K}\\k'\neq k}}g^{(i)}_{k'm}(\hspace{-0.5mm}\boldsymbol{x}_{k}^{(i)})^\text{H}\boldsymbol{x}^{(i)}_{k'}\hspace{-0.5mm}+\hspace{-0.5mm}(\hspace{-0.5mm}\boldsymbol{x}_{k}^{(i)})^\text{H}\boldsymbol{w}^{(i)}_m.
    \label{eq:estimated_2}
\end{align}
Note that the summation term the in last equality captures the effect of PC.

\begin{table}[!t]
\captionsetup{justification=centering, labelsep=newline}
\caption{The amount of overhead in bits per near-RT loop to perform channel estimations}
\label{tb:overhead}
\centering
\begin{tabular}{|c|c|c|c|c|} 
    \hline
    \multirow{2}{*}{Estimation} & \multicolumn{2}{c|}{$T_p=4$} & \multicolumn{2}{c|}{$T_p=8$} \\
    \cline{2-5}
    & $K=24$ & $K=72$ & $K=24$ & $K=72$ \tabularnewline
    \hline
    DU-based & 61,536 & 61,728 & 123,024 & 123,312 \tabularnewline
    \hline
    RU-based & 432 & 1,296 & 648 & 1,944 \tabularnewline
    \hline
\end{tabular}
\end{table}

\subsection{Data Transmission Model}\label{ssec:system_transmission}

For uplink (downlink) data transmission, the estimated channel in~\eqref{eq:estimated_2} is used as a combiner (a precoder), the details of which are given as follows.
For uplink transmission, each user $k$ transmits a data signal $x^{\mathsf{u}}_{k}$.
Then, the received signal $y^{\mathsf{u}}_{m}$ at O-RU $m$ is given by $y^{\mathsf{u}}_{m} = \sum_{k\in\mathcal{K}}g^{(i)}_{km}\sqrt{\rho_{k}}x^{\mathsf{u}}_{k} + w^{\mathsf{u}}_{m}$, where $\rho_{k}$ and $w^{\mathsf{u}}_{m}$ are the transmit power of user $k$ and uplink additive Gaussian noise on O-RU $m$ with variance $\sigma^2_\mathsf{u}$, respectively.
For each user $k\in\mathcal{K}^{\mathsf{RU}}_m$, O-RU $m$ computes $(\widehat{g}^{(i)}_{km})^*y^{\mathsf{u}}_{m}$ and transfers it to the user's serving O-DU.
After collecting the conjugate-multiplied signals from the O-RUs in $\mathcal{M}^{\mathsf{UE}}_k$, the serving O-DU combines them to obtain the data signal $\bar{x}^{\mathsf{u}}_{k}$ expressed as
\begin{equation}
    \bar{x}^{\mathsf{u}}_{k}=\sum_{m\in{\mathcal{M}^{\mathsf{UE}}_k}}(\widehat{g}^{(i)}_{km})^*y^{\mathsf{u}}_{m} = \sum_{m\in{\mathcal{M}^{\mathsf{UE}}_k}}\sum_{k'\in\mathcal{K}}(\widehat{g}^{(i)}_{km})^*g^{(i)}_{k'm}\sqrt{\rho_{k'}}x^{\mathsf{u}}_{k'} + \sum_{m\in{\mathcal{M}^{\mathsf{UE}}_k}}(\widehat{g}^{(i)}_{km})^*w^{\mathsf{u}}_{m}.
    \label{eq:UL_signal}
\end{equation}
Based on~\eqref{eq:UL_signal} and the formulation in \cite{Zhang20}, the effective uplink signal to interference plus noise ratio (SINR) of user $k$ is given by
\begin{equation}
    \hspace{-2mm}\mathsf{SINR}^{\mathsf{u}}_{k}=\frac{\tiny\rho_{k}\left\vert\mathbb{E}\left[\sum_{m\in{\mathcal{M}^{\mathsf{UE}}_k}}(\widehat{g}^{(i)}_{km})^*g^{(i)}_{km}\right]\right\vert^2} {\tiny \displaystyle\sum_{k'\in\mathcal{K}}\rho_{k'}\mathbb{E}\Bigg[\bigg\vert\sum_{m\in\mathcal{M}^{\mathsf{UE}}_k}(\widehat{g}^{(i)}_{km})^*g^{(i)}_{k'm}\bigg\vert^2\Bigg] - \rho_{k}\Bigg\vert\mathbb{E}\Bigg[\sum_{m\in{\mathcal{M}^{\mathsf{UE}}_k}}(\widehat{g}^{(i)}_{km})^*g^{(i)}_{km}\Bigg]\Bigg\vert^2+\sigma^2_\mathsf{u}\sum_{m\in{\mathcal{M}^{\mathsf{UE}}_k}}\mathbb{E}\left[\vert\widehat{g}^{(i)}_{km}\vert^2\right]},
    \label{eq:SINR_UL}
\end{equation}
where the expectation is over the random variables.

For downlink transmission, the data signal $x^{\mathsf{d}}_{k}$ is transmitted by the O-RUs serving user $k$ (i.e., O-RU $m\in\mathcal{M}^{\mathsf{UE}}_k$) after applying the conjugate beamforming expressed as $\bar{x}^{\mathsf{d}}_{km}=(\widehat{g}^{(i)}_{km})^*x^{\mathsf{d}}_{k}/\vert\widehat{g}^{(i)}_{km}\vert$.
The received signal $\bar{y}^{\mathsf{d}}_{k}$ for user $k$ is then given by
\begin{equation}
    \bar{y}^{\mathsf{d}}_{k} = \sum_{k'\in\mathcal{K}}\sum_{m\in\mathcal{M}^{\mathsf{UE}}_{k'}}g^{(i)}_{km}\bar{x}^{\mathsf{d}}_{k'm} + w^{\mathsf{d}}_{k} = \sum_{k'\in\mathcal{K}}\sum_{m\in\mathcal{M}^{\mathsf{UE}}_{k'}}g^{(i)}_{km}(\widehat{g}^{(i)}_{k'm})^*x^{\mathsf{d}}_{k'} + w^{\mathsf{d}}_{k},
    \label{eq:DL_signal}
\end{equation}
where $w^{\mathsf{d}}_{k}$ is the downlink additive noise on user $k$ with variance $\sigma^2_\mathsf{d}$.
Based on~\eqref{eq:DL_signal} and the approach in \cite{Zhang20}, the effective downlink SINR is given by
\begin{equation}
    \mathsf{SINR}^{\mathsf{d}}_{k} = \frac{\left\vert\mathbb{E}\left[\sum_{m\in{\mathcal{M}^{\mathsf{UE}}_k}}g^{(i)}_{km}(\widehat{g}^{(i)}_{km})^*\right]\right\vert^2} {\sum_{k'\in\mathcal{K}}\mathbb{E}\left[\left\vert\sum_{m\in\mathcal{M}^{\mathsf{UE}}_{k'}}g^{(i)}_{km}(\widehat{g}^{(i)}_{k'm})^*\right\vert^2\right] - \left\vert\mathbb{E}\left[\sum_{m\in{\mathcal{M}^{\mathsf{UE}}_k}}g^{(i)}_{km}(\widehat{g}^{(i)}_{km})^*\right]\right\vert^2 + \sigma^2_\mathsf{d}},
    \label{eq:SINR_DL}
\end{equation}
where the expectation is over the random variables.

Based on~\eqref{eq:SINR_UL} and~\eqref{eq:SINR_DL}, the achievable uplink and downlink spectral efficiencies (SEs) for user $k$ are computed as $R^{\mathsf{u}}_{k} = \log_2(1+\mathsf{SINR}^{\mathsf{u}}_{k})$ and $R^{\mathsf{d}}_{k} = \log_2(1+\mathsf{SINR}^{\mathsf{d}}_{k})$, respectively.
Note that these SE metrics can be used to quantify the uplink/downlink data transmission performance~\cite{Ngo17,Zhang20}.
Since the SINR expressions contain the estimated channel term $\widehat{g}^{(i)}_{km}$, the performance is directly impacted by the channel estimation performance our work focuses to improve.

\subsection{Problem Formulation}\label{ssec:system_formulation}

We use MSE of the channel estimation described  in Sec.~\ref{ssec:system_estimation} for our PA performance metric.
For user $k$ served by the O-RUs in $\mathcal{M}^\mathsf{UE}_k$, we define the MSE of the channel estimate in~\eqref{eq:estimated_2} as
\begin{align}
    \mathsf{MSE}^{(i)}_k   &=\mathbb{E}\bigg[\sum_{m\in\mathcal{M}^\mathsf{UE}_k}\Big\vert\widehat{g}^{(i)}_{km}-g^{(i)}_{km}\Big\vert^2\bigg] =\hspace{-2mm}\sum_{m\in\mathcal{M}^\mathsf{UE}_k}\mathbb{E}\left[\Big\vert\widehat{g}^{(i)}_{km}-g^{(i)}_{km}\Big\vert^2\right] \nonumber \\ 
    &\hspace{-10mm}=\hspace{-2mm}\sum_{m\in\mathcal{M}^\mathsf{UE}_k}\mathbb{E}\Bigg[\Big\vert\sum_{\substack{k'\in\mathcal{K}\\k'\neq k}}g^{(i)}_{k'm}(\boldsymbol{x}_{k}^{(i)})^\text{H}\boldsymbol{x}^{(i)}_{k'} +(\boldsymbol{x}_{k}^{(i)})^\text{H}\boldsymbol{w}^{(i)}_m\Big\vert^2\Bigg] =\hspace{-2mm}\sum_{m\in\mathcal{M}^\mathsf{UE}_k}\sum_{\substack{k'\in\mathcal{K}\\k'\neq k}}\beta^{(i)}_{k'm}\Big|(\boldsymbol{x}_{k}^{(i)})^\text{H}\boldsymbol{x}^{(i)}_{k'}\Big|^2 +\sigma^2,\label{eq:MSE_4}
\end{align}
where the expectation is taken over the channel and noise.
The third equality holds as we substitute $\widehat{g}^{(i)}_{km}$ with~\eqref{eq:estimated_2}.
Next, the last equality holds since (i) $g^{(i)}_{km}$ and $\boldsymbol{w}^{(i)}_{m}$ are i.i.d. across $k$ and $m$ and (ii) $\mathbb{E}[|g^{(i)}_{km}|^2]=\beta^{(i)}_{km}$ and $\mathbb{E}[\|\boldsymbol{w}^{(i)}_{m}\|^2_2]=\sigma^2$.
From~\eqref{eq:MSE_4}, we see that the MSE is directly proportional to the interference caused by PC, and thus can be used as an effective metric to quantify the PA performance.

Since our system involves $U$ agents, each of which handles the PA of user $k\in\mathcal{K}^\mathsf{DU}_u$, we can formulate the PA optimization problem for agent $u$ as
\begin{align}
    (\boldsymbol{\mathcal{P}}_u):\;\;&\min_{\{\boldsymbol{x}^{(i)}_k\}_{k\in\mathcal{K}^\mathsf{DU}_u}} \sum_{k\in\mathcal{K}}\mathsf{MSE}^{(i)}_k \label{eq:prob_obj} \\
    &\text{s.t.}\;\;\boldsymbol{x}^{(i)}_k\in\mathcal{T}^{(i)}_{u}, \;\; \forall k \in \mathcal{K}^\mathsf{DU}_u, \label{eq:prob_const1} \\
    &\hspace{6mm}\|\boldsymbol{\phi}^{(i)}_{u,t}\|^2_2 = 1, \left(\boldsymbol{\phi}_{u,t}^{(i)}\right)^{\hspace{-0.7mm}\text{H}}\hspace{-1mm}\boldsymbol{\phi}^{(i)}_{u,t'} = 0 \text{ if } t \neq t', \;\; \forall t,t'=1,2,\ldots,T_p.
\end{align}
If $\beta_{km}^{(i)}$, $\forall k,m$ is known, one can directly evaluate $\sum_{k\in\mathcal{K}}\mathsf{MSE}^{(i)}_k$ using~\eqref{eq:MSE_4} and solve $\boldsymbol{\mathcal{P}}_u$ using an existing PA algorithm (e.g., the previous works~\cite{Ngo17,Attarifar18,Sabbagh18,Chen21_1,Liu20_1,Liu20_2,Buzzi21}).
However, in large-scale systems, such prior knowledge is often not available, and one can no longer evaluate the objective function in a straightforward manner.
Suppose the knowledge is somehow available for the MSE to be evaluated, but some of these algorithms (e.g., PAs using the  Tabu-search~\cite{Liu20_1} and Hungarian algorithm~\cite{Buzzi21} having the complexities of $\mathcal{O}(N_\mathsf{tabu}K^2M)$ and $\mathcal{O}(KT_p^3)$, respectively) still cannot be considered as the complexity becomes prohibitive for a large number of users.
To address both issues, we solve $\boldsymbol{\mathcal{P}_u}$ via a distributed learning framework, details of which are given in Sec.~\ref{sec:methodology}.
The decentralization imposed in this work allows our PA scheme to be much more scalable.

\section{Scalable Learning-based Pilot Assignment Scheme for O-RAN CFmMIMO}\label{sec:methodology}

In this section, we first describe how our proposed PA scheme is framed in O-RAN (Sec.~\ref{ssec:methodology_ORAN}). 
Next, after providing preliminaries on MA-DRL (Sec.~\ref{ssec:methodology_prelim}), we design a Markov game model perceiving our PA problem (Sec.~\ref{ssec:methodology_model}), and show that the action selection in our learning framework corresponds to minimizing the PC (Theorem~\ref{prop:prop_1}).
Finally, we provide implementation details for our DRL-based PA scheme (Sec.~\ref{ssec:methodology_framework}) and iterative CS algorithm (Sec.~\ref{ssec:methodology_codebook}). 

\subsection{Pilot Assignment Framework in O-RAN Architecture}\label{ssec:methodology_ORAN}

\begin{figure}[!t]
    \centering
    \includegraphics[width=1\linewidth]{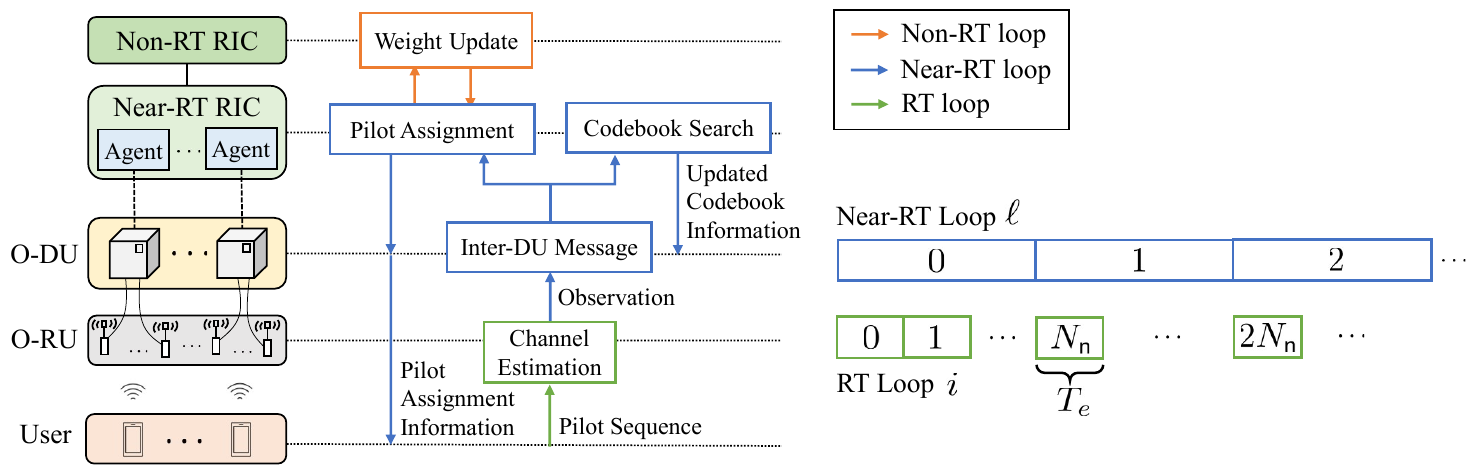}
    \caption{A block diagram of the proposed PA scheme.}
    \label{fig:PA_structure}
\end{figure}

Our learning-based PA scheme for CFmMIMO is designed based on the  O-RAN architecture defined in Sec.~\ref{ssec:system_cellfree}.
Its conceptual block diagram is illustrated in Fig.~\ref{fig:PA_structure}.
Here the PA is conducted under three different O-RAN control loops which have been described earlier in Fig.~\ref{fig:ORAN_architecture}.

\subsubsection{RT loop} We assume that a single round of channel estimation steps described in Sec.~\ref{ssec:system_estimation} takes place in each RT loop.
Hence, we denote the index of each RT loop using the same notation used for indexing the channel estimation instance.
In each RT loop $i$, users transmit their assigned pilots, and the O-RU $m$ completes the channel estimation to obtain $\widehat{g}^{(i)}_{km}$ for $k \in \mathcal{K}^\mathsf{RU}_m$.

\subsubsection{Near-RT loop} Near-RT loop occurs once in every $N_\mathsf{n}$ RT loops. During each near-RT loop, O-DU $u$ collects observation data, which we describe later in Sec.~\ref{ssec:methodology_model}, from the O-RUs in $\mathcal{M}^\mathsf{DU}_u$ and transfers it to the near-RT RIC to be used for learning.
At the same time, each agent $u$ in the near-RT RIC conducts PA on the users in $\mathcal{K}^\mathsf{DU}_u$.
We use $\ell = 0,1,\ldots,\lfloor\frac{N}{N_\mathsf{n}}\rfloor$ to denote the index of near-RT loop, thus, $\ell$-th near-RT loop occurs during the $N_\mathsf{n}\ell$-th RT loop (or the $N_\mathsf{n}\ell$-th channel estimation instance).
The relationship between $i$ and $\ell$ is visualized in Fig.~\ref{fig:PA_structure}.

To further improve our PA performance, two acceleration techniques are introduced:
\begin{itemize}
    \item \emph{Inter-DU message passing}: We consider inter-DU message passing which occurs at each near-RT loop.
    The inter-DU connection is essential for fully realizing user-centric RU clusters in decentralized CFmMIMO~\cite{Ranjbar22}, and we exploit this feature to improve our PA performance.
    With inter-DU messages, we aim to reinforce the data observed by the local group of O-RUs (i.e., O-RUs of $\mathcal{M}^\mathsf{DU}_u$).
    The details on inter-DU message passing are provided in Sec.~\ref{ssec:methodology_framework}.
    \item \emph{Codebook searching}: We leverage the decentralization of our system and develop a CS algorithm that operates jointly with our PA scheme.
    We adopt the idea of quasi-orthogonal codebooks~\cite{Wang20,Chowdhury22} to be used across the agents.
    In multi-cell systems, where each cell conducts its own PA to the serving users, using~non-identical orthogonal codebooks across the cells has shown improved system performance~\cite{Wang20,Chowdhury22}.~Inspired by this, we rotate the codebook of each agent in an iterative manner to find the codebook orientation that yields the minimum MSE of channel estimation.
    The detailed steps of our CS scheme are provided in Sec.~\ref{ssec:methodology_codebook}.
\end{itemize}

\subsubsection{Non-RT loop} The non-RT loop is utilized to handle time-insensitive tasks. In our PA scheme, the update of the learning parameters for near-RT RIC occurs over this loop.
Here, the non-RT loop occurs once in every $N_\mathsf{non}$ RT loops, and we denote $q = 0,1,\ldots,\lfloor\frac{N}{N_\mathsf{non}}\rfloor$ as the non-RT loop index.
As described in Fig.~\ref{fig:ORAN_architecture}, a near-RT loop duration can be as short as 10 ms while the shortest duration for non-RT loop is a second~\cite{ORAN}. Hence, we assume $N_\mathsf{non} \gg N_\mathsf{n}$.

\subsection{Preliminaries on Multi-agent Deep Reinforcement Learning}\label{ssec:methodology_prelim}

MA-DRL addresses scenarios where multiple agents perform simultaneous decision-making based on a Markov game model~\cite{Qu22}.
For our decentralized PA problem, we define MA-DRL using a tuple $(\{\boldsymbol{S}^{(\ell)}_u\}_{u\in\mathcal{U}}$, $\{\boldsymbol{a}^{(\ell)}_u\}_{u\in\mathcal{U}}$, $\{r^{(\ell)}_u\}_{u\in\mathcal{U}})$, where $\boldsymbol{S}^{(\ell)}_u$, $\boldsymbol{a}^{(\ell)}_u$, and $r^{(\ell)}_u$ are respectively the \emph{state}, \emph{action}, and \emph{reward} of the agent $u$ during the $\ell$-th near-RT loop.
For each loop $\ell$, agent $u$ with a state $\boldsymbol{S}^{(\ell)}_u$ makes an action $\boldsymbol{a}^{(\ell)}_u$ to interact with the environment.
Subsequently, the agent makes an observation and computes a reward $r^{(\ell)}_u$ which helps to find the next state $\boldsymbol{S}^{(\ell+1)}_u$.

In the non-RT loop, once an agent has completed multiple interactions with the environment, its policy on action selection for a given state is optimized by updating the weights of its respective deep neural network (DNN).
Here the action is determined based on the Q-value~\cite{RL} denoted by $Q(\boldsymbol{S}^{(\ell)}_u,\boldsymbol{a}^{(\ell)}_u)$.
The Q-value quantifies the quality of an agent's action for a given state.
Thus, it is important for the agent to obtain accurate Q-values to make correct decisions.
In DRL, these Q-values are computed via a DNN, the weights of which are trained with experiences so that a correct (i.e., Q-value-maximizing) action can be selected upon each decision-making.

Now, in perceiving our PA task as a multi-agent learning problem, there are two conditions we need to consider~\cite{Feriani21}.
First, multiple agents making independent decisions simultaneously implies the environment is never seen as stationary to an action of a single agent.
Second, due to the decentralized architecture, each agent only obtains a part of the observation available from the entire environment.
Due to these conditions, in multi-agent learning, careful design of the Markov game model is crucial for achieving performance comparable to centralized learning.

\subsection{Markov Game Model for Decentralized Pilot Assignment}\label{ssec:methodology_model}

In our O-RAN CFmMIMO setting, channel estimation is repeated for every RT loop $i$, forming a periodic interaction with the environment.
The near-RT PA corresponds to action selection that affects the environment and resulting observation.
Based on this, we formally define each component of the tuple presented in Sec.~\ref{ssec:methodology_prelim} to perceive our PA task as a Markov game model.

\subsubsection{States} To represent the PA status of agent $u$ on users in $\mathcal{K}^\mathsf{DU}_u$ at the start of near-RT loop~$\ell$, we define the state as $\boldsymbol{S}_u^{(\ell)} = \boldsymbol{\Phi}_u^{(\ell)}$ which is a $|\mathcal{K}^\mathsf{DU}_u|\times T_p$ sized matrix where
\begin{equation}
    [\boldsymbol{\Phi}_u^{(\ell)}]_{k,t} =
    \begin{cases}
        1 & \text{if } \boldsymbol{x}^{(N_\mathsf{n}\ell)}_k = \boldsymbol{\phi}^{(N_\mathsf{n}\ell)}_{u,t}, \\
        0 & \text{otherwise}.
    \end{cases}
    \label{eq:state_entry}
\end{equation}
As discussed previously, PC occurs when users share a pilot, and this can be indicated by the \emph{ones} in each column of $\boldsymbol{\Phi}_u^{(\ell)}$.
Hence, $\boldsymbol{\Phi}_u^{(\ell)}$ can become an effective means to represent the condition of PA for each agent, and we aim to have the agents accurately perceive the relationship between their PA (i.e., their actions) and the resulting PC.

\subsubsection{Actions} We consider sequential updates on the pilots, where the pilot of only a single user is changed with every action.
If we consider actions that assign pilots to all $|\mathcal{K}^\mathsf{DU}_u|$ users at once, this would lead our action space to take $T_p^{|\mathcal{K}^\mathsf{DU}_u|}$ possible combinations and suffer from the ``curse of dimensionality''.
We hence define actions as an ordered pair indicating the user of interest and the pilot to be assigned, respectively.
The action of agent $u$ at near-RT PA $\ell$ is formally defined as
$\boldsymbol{a}_u^{(\ell)}=(k,t)$, where $k\in\mathcal{K}^\mathsf{DU}_u$ and $t\in\{1, 2, \ldots,T_p\}$. With this setting, there are total $|\mathcal{K}^\mathsf{DU}_u|T_p$ possible actions for agent $u$ to take, resulting in a more computationally scalable action space.

\subsubsection{Rewards} We propose to compute the reward of each agent $u$ on the $\ell$-th near-RT PA based on the average sum-power of the channel estimates obtained by the O-RUs.
Note that, for each action (i.e., near-RT PA) taken by an agent, $N_\mathsf{n}$ channel estimations are conducted by O-RU $m$ to acquire a set of $\widehat{\boldsymbol{g}}^{(i)}_{m}$ for $N_\mathsf{n}\ell \leq i < N_\mathsf{n}(\ell+1)$.
Using this information, the O-RU $m$ computes
\begin{equation}
    p^{(\ell)}_{km} = \frac{1}{N_\mathsf{n}}\sum_{n=0}^{N_\mathsf{n}-1}\left|\widehat{g}^{(N_\mathsf{n}\ell+n)}_{km}\right|^2
    \label{eq:sum_power_km}
\end{equation}
on user $k\in\mathcal{K}^\mathsf{RU}_m$ during the near-RT loop $\ell$ and sends it to the corresponding O-DU.
At the end of this transfer, O-DU $u$ collects different sets of $p^{(\ell)}_{km}$ from each O-RU $m \in \mathcal{M}^\mathsf{DU}_u$ (i.e., $\{\{p^{(\ell)}_{km}\}_{k\in\mathcal{K}^\mathsf{RU}_m}\}_{m\in\mathcal{M}^\mathsf{DU}_u}$).
In decentralized PA, each agent $u \in \mathcal{U}$ is responsible for a disjoint subset of $K$ users, and it is desirable for the agent to have access to $p^{(\ell)}_{km}$ from all O-RUs associated with the users (i.e., $\{\{p^{(\ell)}_{km}\}_{m\in\mathcal{M}^\mathsf{UE}_k}\}_{k\in\mathcal{K}^\mathsf{DU}_u}$).
However, as each O-DU $u$ is only connected to O-RUs of $\mathcal{M}^\mathsf{DU}_u$, $\{\{p^{(\ell)}_{km}\}_{m\in\mathcal{M}^\mathsf{UE}_k\cap\mathcal{M}^\mathsf{DU}_u}\}_{k\in\mathcal{K}^\mathsf{DU}_u}$ only gets collected by the agent.
Hence, O-DU $u$ ends up computing the observation data to be transferred to the agent $u$ as $\bar{p}^{(\ell)}_{u} = \sum_{k\in\mathcal{K}^\mathsf{DU}_u}\sum_{m\in\mathcal{M}^\mathsf{UE}_k\cap\mathcal{M}^\mathsf{DU}_u} p^{(\ell)}_{km}$.

Note that the rest of information required by agent $u$ (i.e., $\{\{p^{(\ell)}_{km}\}_{m\in\mathcal{M}^\mathsf{UE}_k\setminus\mathcal{M}^\mathsf{DU}_u}\}_{k\in\mathcal{K}^\mathsf{DU}_u}$) has been collected by other O-DUs.
As mentioned earlier in Sec.~\ref{ssec:methodology_ORAN}, since we consider inter-DU messages, this information can be transferred to each corresponding O-DU.
Then, each O-DU $u$ can now compute the reinforced observation data which is expressed as
\begin{align}
    \widetilde{p}^{(\ell)}_{u} &= \bar{p}^{(\ell)}_{u} + \sum_{k\in\mathcal{K}^\mathsf{DU}_u}\sum_{m\in\mathcal{M}^\mathsf{UE}_k\setminus\mathcal{M}^\mathsf{DU}_u} p^{(\ell)}_{km} = \sum_{k\in\mathcal{K}^\mathsf{DU}_u}\sum_{m\in\mathcal{M}^\mathsf{UE}_k}p^{(\ell)}_{km}.
    \label{eq:pu_tilde}
\end{align}

The observation data computed by O-DU $u$ in~\eqref{eq:pu_tilde} is transferred to agent $u$ via a backhaul,~and the reward for agent $u$ at near-RT loop $\ell$ is subsequently computed using the mapping function
\begin{equation}
    r^{(\ell)}_u(p) = (p_\text{max}-p)/(p_\text{max}-p_\text{min}),
    \label{eq:reward}
\end{equation}
where $p=\widetilde{p}^{(\ell)}_{u}$ by the availability of inter-DU message.
The mapping function~\eqref{eq:reward} converts the observation data into a reward range such that lower values of $p$ are rewarded higher.
Here $[p_\text{min},p_\text{max}]$ is the range of observation data, which we assume is set by the non-RT RIC.

We now show that the learning via our Markov model leads to taking an action that minimizes the degree of PC.
The basic mechanism of learning we utilize is that, for each given state $\boldsymbol{S}_u$, we want the agent $u$ to select the action that maximizes its Q-value~\cite{RL}, i.e.,
\begin{equation}
    \boldsymbol{a}^\star_{u} = \argmax_{\boldsymbol{a}_u\in\mathcal{A}_u} Q(\boldsymbol{S}_u,\boldsymbol{a}_u),
    \label{eq:action_selection}
\end{equation}
where $\mathcal{A}_u$ is the set of all possible actions for agent $u$.
The training in DRL is done by updating the network weights via regression toward the experiences obtained.
The Q-value, which is the numerical output of the trained network, is then expected to follow the average of these experiences,
i.e., the Q-value is updated through training to yield $Q(\boldsymbol{S}_u,\boldsymbol{a}_u) = \mathbb{E}[r_u(p)|(\boldsymbol{S}_u,\boldsymbol{a}_u)].$

For each near-RT loop $\ell$, the following theorem shows that, with inter-DU message passing, the action selected via~\eqref{eq:action_selection} is the best action in terms of minimizing the degree of local PC.
\begin{theorem}
    With $\widetilde{p}^{(\ell)}_u$ available, for a given state $\boldsymbol{S}^{(\ell)}_u$, taking the action $\boldsymbol{a}^{(\ell)}_u$ which satisfies~\eqref{eq:action_selection} is equivalent to finding the action that minimizes the degree of pilot contamination occurring on local users in $\mathcal{K}^\mathsf{DU}_u$ during the near-RT loop $\ell$, which is expressed as
    \begin{equation}
        \sum_{k\in\mathcal{K}^\mathsf{DU}_u} \sum_{m\in\mathcal{M}^\mathsf{UE}_k} \sum_{n=0}^{N_\mathsf{n}-1} \sum_{k'\in\mathcal{K},k'\neq k}\beta^{(N_\mathsf{n}\ell+n)}_{k'm}\Big|(\boldsymbol{x}_{k}^{(N_\mathsf{n}\ell)})^\text{H}\boldsymbol{x}^{(N_\mathsf{n}\ell)}_{k'}\Big|^2.\label{eq:prop_1}
    \end{equation}
    \label{prop:prop_1}
\end{theorem}
\vspace{-5mm}
\begin{proof}
    First, in terms of the parameters defined in our model, we find the expected reward at near-RT loop $\ell$ for a given state-action pair $(\boldsymbol{S}^{(\ell)}_u,\boldsymbol{a}^{(\ell)}_u)$, which is expressed as 
    \begin{equation}
        \mathbb{E}[r_u^{(\ell)}(\widetilde{p}^{(\ell)}_u)|(\boldsymbol{S}^{(\ell)}_u,\boldsymbol{a}^{(\ell)}_u)] = \frac{p_\text{max}-\mathbb{E}[\widetilde{p}^{(\ell)}_u]}{p_\text{max}-p_\text{min}},
        \label{eq:expected_reward}
    \end{equation}
    where the equality holds from~\eqref{eq:reward}.
    Recalling~\eqref{eq:action_selection}, the learning conducted at each agent $u$ aims to find the action achieving the maximum Q-value $Q(\boldsymbol{S}_u,\boldsymbol{a}_u)$, which we discussed to yield $\mathbb{E}[r_u(p)|(\boldsymbol{S}_u,\boldsymbol{a}_u)]$.
    Thus, the action selection mechanism of agent $u$ can be expressed as
    \begin{equation}
        \boldsymbol{a}_{u}^{(\ell)} =\argmax_{\boldsymbol{a}_u\in\mathcal{A}_u}\mathbb{E}[r_u^{(\ell)}(\widetilde{p}^{(\ell)}_u)|(\boldsymbol{S}^{(\ell)}_u,\boldsymbol{a}_u)].
        \label{eq:action_selection2}
    \end{equation}
    Now combining~\eqref{eq:expected_reward} and~\eqref{eq:action_selection2}, we can say that
    \begin{align}
        \hspace{-4mm}\boldsymbol{a}_{u}^{(\ell)}=\argmin_{\boldsymbol{a}_u\in\mathcal{A}_u}\;\sum_{k\in\mathcal{K}^\mathsf{DU}_u}\sum_{m\in\mathcal{M}^\mathsf{UE}_k}\mathbb{E}[p^{(\ell)}_{km}] =\argmin_{\boldsymbol{a}_u\in\mathcal{A}_u}\frac{1}{N_\mathsf{n}}\sum_{k\in\mathcal{K}^\mathsf{DU}_u}\sum_{m\in\mathcal{M}^\mathsf{UE}_k}\sum_{n=0}^{N_\mathsf{n}-1}\mathbb{E}\left[\left|\widehat{g}^{(N_\mathsf{n}\ell+n)}_{km}\right|^2\right], \label{eq:action_selection3_2}
    \end{align}
    where the first and second equalities are obtained using \eqref{eq:pu_tilde} and \eqref{eq:sum_power_km}, respectively.
    Now, for $n = 0, 1, \dots, N_\mathsf{n}-1$, using \eqref{eq:estimated_2} we have
    \begin{align}
        \mathbb{E}\left[\left|\widehat{g}^{(N_\mathsf{n}\ell+n)}_{km}\right|^2\right]&=\mathbb{E}\bigg[\Big|g^{(N_\mathsf{n}\ell+n)}_{km}\Big|^2\bigg]+\sum_{\substack{k'\in\mathcal{K}\\k'\neq k}}\mathbb{E}\bigg[\Big|g^{(N_\mathsf{n}\ell+n)}_{k'm}(\boldsymbol{x}_{k}^{(N_\mathsf{n}\ell+n)})^\text{H}\boldsymbol{x}^{(N_\mathsf{n}\ell+n)}_{k'}\Big|^2\bigg] \nonumber \\
        &\hspace{10mm}+\mathbb{E}\bigg[\Big|(\boldsymbol{x}_{k}^{(N_\mathsf{n}\ell+n)})^\text{H}\boldsymbol{w}^{(N_\mathsf{n}\ell+n)}_m\Big|^2\bigg] =\beta^{(N_\mathsf{n}\ell+n)}_{km}+\xi^{(\ell,n)}_{km}+\sigma^2, \label{eq:expected_ch_pow5}
    \end{align}
    where $\xi^{(\ell,n)}_{km}=\sum_{\substack{k'\in\mathcal{K} \\ k'\neq k}}\beta^{(N_\mathsf{n}\ell+n)}_{k'm}\big|(\boldsymbol{x}_{k}^{(N_\mathsf{n}\ell+n)})^\text{H}\boldsymbol{x}^{(N_\mathsf{n}\ell+n)}_{k'}\big|^2$ reflects the PC discussed in Sec.~\ref{ssec:system_formulation}.
    By the definition of $\widehat{g}^{(i)}_{km}$ in~\eqref{eq:estimated_2}, taking the expectation of $|\widehat{g}^{(i)}_{km}|^2$ leaves only the autocorrelation terms for $\widehat{g}^{(i)}_{km}$ and $\boldsymbol{w}^{(i)}_m$, corresponding to $\beta_{km}^{(N_\mathsf{n}\ell+n)} = \mathbb{E}[|g_{km}^{(N_\mathsf{n}\ell+n)}|^2]$ and $\sigma^2 = \mathbb{E}[|(\boldsymbol{x}_{k}^{(N_\mathsf{n}\ell+n)})^\text{H}\boldsymbol{w}^{(N_\mathsf{n}\ell+n)}_m|^2]$ in~\eqref{eq:expected_ch_pow5}. 
    This is because the channel and noise are assumed uncorrelated across $k$ and $m$.

    Now, since (i) $\xi^{(\ell,n)}_{km}$ is the only term that is impacted by action $\boldsymbol{a}_u$, i.e., $\beta^{(N_\mathsf{n}\ell+n)}_{km}$ and $\sigma^2$ in~\eqref{eq:expected_ch_pow5} are independent from PA and (ii) $\boldsymbol{x}_k^{(i)}$ only changes once every $N_\mathsf{n}$ RT loops, i.e., $\boldsymbol{x}_k^{(N_\mathsf{n}\ell+n)}$ is fixed for $n=0,1,\ldots,N_\mathsf{n}-1$, by ignoring $\frac{1}{N_\mathsf{n}}$ as a scaling factor, \eqref{eq:action_selection3_2} is equivalent to
    \begin{align}
        \boldsymbol{a}_{u}^{(\ell)}=\argmin_{\boldsymbol{a}_u\in\mathcal{A}_u}\;& \sum_{k\in\mathcal{K}^\mathsf{DU}_u} \sum_{m\in\mathcal{M}^\mathsf{UE}_k} \sum_{n=0}^{N_\mathsf{n}-1} \sum_{k'\in\mathcal{K},k'\neq k}\beta^{(N_\mathsf{n}\ell+n)}_{k'm}\Big|(\boldsymbol{x}_{k}^{(N_\mathsf{n}\ell)})^\text{H}\boldsymbol{x}^{(N_\mathsf{n}\ell)}_{k'}\Big|^2,
    \end{align}
    which represents the degree of PC at near-RT loop $\ell$ over the users in $\mathcal{K}^\mathsf{DU}_u$.
\end{proof}

From Theorem~\ref{prop:prop_1}, we conclude that learning based on our Markov games model is equivalent to performing the pilot update which minimizes the interference due to PC at each near-RT PA.
According to~\eqref{eq:prop_1}, the PA made at each near-RT loop $\ell$ couples with the pathloss occurring over the corresponding $N_\mathsf{n}$ RT loops.
Since we do not assume prior knowledge on the pathloss $\beta^{(i)}_{km}$, we cannot evaluate the exact MSE.
However, through the reward we define and the learning mechanism of DRL, we can still design our PA scheme such that the MSE performance is improved over time.
For static scenarios, where $\beta^{(i)}_{km}$ is constant over $i$, all the actions taken over near-RT loops (i.e., the entire series of successive pilot updates) are contributing to look for a single optimal PA solution that minimizes the sum-MSE.
On the other hand, for mobile scenarios, each action is led to focus on minimizing the sum-MSE resulted from the current channel statistics by leveraging the past information.
Our PA scheme is designed to cope with time varying small-scale and large-scale fading factors upon continuous training. 

\begin{figure}[t]
    \centering
    \includegraphics[width=0.7\linewidth]{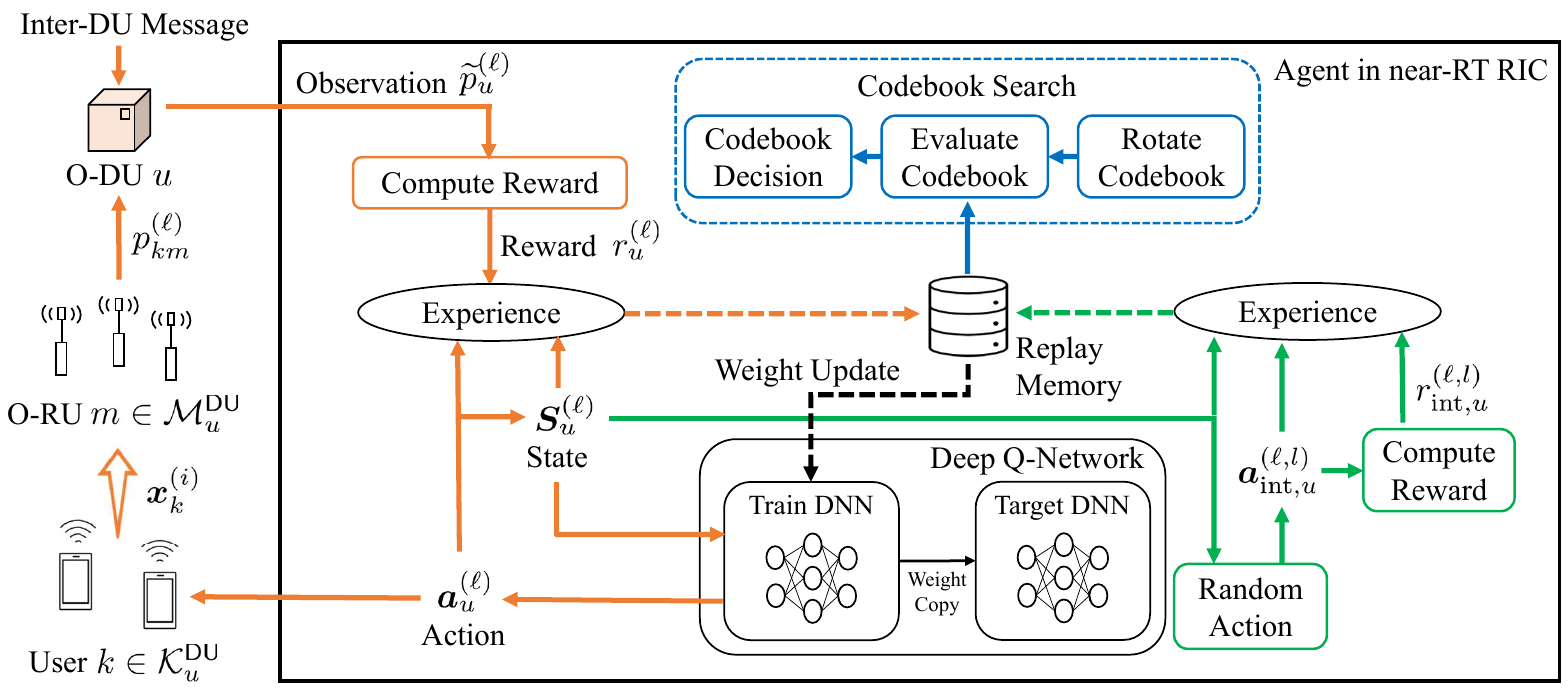}
    \caption{A block diagram overview of our PA scheme, consisting of non-RT DNN training and near-RT PA updates.}
    \label{fig:DRL_architecture}
\end{figure}

\subsection{MA-DRL-based Pilot Assignment Scheme}\label{ssec:methodology_framework}

Given the setting in the previous subsections, we describe our PA scheme in detail using the MA-DRL framework to find the solution to our decentralized PA problem.
Our PA scheme trains learning algorithms to perform sequential pilot updates that aim to reduce the sum average power of channel estimates across the users.
As shown by Theorem~\ref{prop:prop_1}, minimizing the sum average power of channel estimates is equivalent to minimizing the overall MSE of channel estimation since both expressions share the PC term that is directly impacted by PA.
Through sequential updates, our algorithm follows a greedy search framework that significantly reduces the dimension of the action space for a more practical MSE minimization.
We incorporate MA-DRL via the deep Q-network (DQN) that utilizes neural network layers for approximating Q-values.
An individual DQN is implemented at each agent in the near-RT RIC for distributed learning.
Fig.~\ref{fig:DRL_architecture} provides an overview of our methodology, which is also outlined in Alg.~\ref{alg:PA-DRL}.
We detail each of the steps in the following:

\textbf{Near-RT PA:} At $\ell=0$, each agent $u$ randomly assigns one of the $T_p$ sequences in $\mathcal{T}^{(0)}_u$ to its associated users in $\mathcal{K}^\mathsf{DU}_u$, from which the state $\boldsymbol{S}^{(0)}_u$ is generated. 
For each subsequent near-RT loop $\ell$, the agent $u$ takes an action $\boldsymbol{a}^{(\ell)}_u$ via an $\epsilon$-greedy method~\cite{RL} to update one user's pilot sequence and obtain a new state $\boldsymbol{S}^{(\ell+1)}_u$.
If the agent decides to take its action based on Q-values, the state $\boldsymbol{S}^{(\ell)}_u$ is used as a $|\mathcal{K}^\mathsf{DU}_u|\times T_p$ input to the DQN, which outputs the Q-value vector of size $|\mathcal{K}^\mathsf{DU}_u|T_p$.
The action with the highest Q-value is then selected.
Since $N_\mathsf{n}$ RT channel estimations occur during a single loop of near-RT PA, each O-DU $u$ collects the necessary information, i.e., $\{p^{(\ell)}_{km}\}_{k\in\mathcal{K}^\mathsf{RU}_m}$, from the O-RUs in $\mathcal{M}^\mathsf{DU}_u$ and computes $\widetilde{p}_{u}^{(\ell)}$ with the aid of inter-DU message passing.
The O-DU transfers $\widetilde{p}_{u}^{(\ell)}$ to its agent in the near-RT RIC, which computes the reward $r^{(\ell)}_u(p)$ and stores an experience tuple $(\boldsymbol{S}_u^{(\ell)}, \boldsymbol{a}_u^{(\ell)}, r_u^{(\ell)}(p), \boldsymbol{S}_u^{(\ell+1)})$ in a \emph{replay memory} of size $D_\mathsf{m}$.

{\footnotesize
\begin{algorithm}[!t]
    \caption{Proposed Pilot Assignment Scheme}
    \label{alg:PA-DRL}
    {\small
    \textbf{Input:} Pilot length $T_p$, number of RT loops $N$, number of RT loops per near-RT loop $N_\mathsf{n}$, number of internal loops $L$, set of users managed by O-DU $u$ $\mathcal{K}^{\mathsf{DU}}_u$, set of O-RUs managed by O-DU $u$ $\mathcal{M}^\mathsf{DU}_u$, set of users served by O-RU $m$ $\mathcal{K}^{\mathsf{RU}}_m$, set of O-RUs serving the user $k$ $\mathcal{M}^\mathsf{UE}_k$, training period, update period \\
    Initialize near-RT loop index $\ell = 0$; randomize the parameter vectors $\boldsymbol{\theta}^\mathsf{tr}_u$ and $\boldsymbol{\theta}^\mathsf{ta}_u$ \\
    Generate codebook $\mathcal{T}^{(N_\mathsf{n}\ell)}_u$; randomly assign $\{\boldsymbol{\phi}^{(N_\mathsf{n}\ell)}_k\}_{k\in\mathcal{K}^\textsf{DU}_u}$ \\
    \For{$\ell = 0$ to $N$}{
        Compute $\boldsymbol{S}^{(\ell)}_u$ using~\eqref{eq:state_entry} \\
        \If{$\ell > 0$}{
            Compute $r^{(\ell-1)}_u(\widetilde{p}_{u}^{(\ell-1)})$ using~\eqref{eq:reward}; store $(\boldsymbol{S}_u^{(\ell-1)}, \boldsymbol{a}_u^{(\ell-1)}, r_u^{(\ell-1)}(\widetilde{p}_{u}^{(\ell-1)}), \boldsymbol{S}_u^{(\ell)})$ in the memory \\
            \For{$l=0$ to $L-1$}{
                Select $\boldsymbol{a}^{(\ell,l)}_{\text{int},u}$ randomly; compute $\boldsymbol{S}^{(\ell,l)}_{\text{int},u}$  using~\eqref{eq:state_entry}; compute $r^{(\ell,l)}_{\text{int},u}$ using~\eqref{eq:int_reward} \\
                Store $(\boldsymbol{S}_{u}^{(\ell)}, \boldsymbol{a}_{\text{int},u}^{(\ell,l)}, r_{\text{int},u}^{(\ell,l)}, \boldsymbol{S}_{\text{int},u}^{(\ell,l)})$ in the memory
            }
        }
        \textbf{if} $\epsilon$-greedy \textbf{then} select $\boldsymbol{a}^{(\ell)}_u$ randomly \textbf{else} $\boldsymbol{a}^{(\ell)}_u=\argmax_{\boldsymbol{a}_u}Q_{\boldsymbol{\theta}^\mathsf{tr}_u}(\boldsymbol{S}^{(\ell)}_u,\boldsymbol{a}_u)$ \\        
        Update the PA according to $\boldsymbol{a}^{(\ell)}_u$ \\
        \For{$i=0$ \textrm{to} $N_\mathsf{n}-1$}{
            User $k\in\mathcal{K}^\textsf{DU}_u$ transmits $\boldsymbol{\phi}^{(N_\mathsf{n}\ell+i)}_k$; O-RU $m\in\mathcal{M}^\mathsf{DU}_u$ estimates $\{\widehat{g}^{(i)}_{km}\}_{k\in\mathcal{K}^\mathsf{RU}_m}$ using~\eqref{eq:estimated_2}
        }
        \textbf{if} $\text{mod}(\ell,\text{training period})=0$ \textbf{then} generate a batch from the memory and
            train $\boldsymbol{\theta}^\mathsf{tr}_u$ via SGD on~\eqref{eq:loss} \\
        \textbf{if} $\text{mod}(\ell,\text{update period})=0$ \textbf{then} set $\boldsymbol{\theta}^\mathsf{ta}_u$ = $\boldsymbol{\theta}^\mathsf{tr}_u$
    }
    \textbf{Output:} Updated pilot sequences $\{\boldsymbol{\phi}^{(N)}_k\}_{k\in\mathcal{K}^\textsf{DU}_u}$
    }
\end{algorithm}
}

\textbf{Non-RT DNN Training:} The learning of each agent $u$ is carried out by two DNNs called the \emph{train} and \emph{target} networks~\cite{Ge20,Li21}, where their network parameter vectors are denoted by $\boldsymbol{\theta}^\mathsf{tr}_u$ and $\boldsymbol{\theta}^\mathsf{ta}_u$, respectively.
Once enough experiences have been collected in the memory, a \emph{mini-batch} of size $D_\mathsf{b}$ is randomly selected from the memory and used to update $\boldsymbol{\theta}^\mathsf{tr}_u$ minimizing the loss:
\begin{equation}
    L(\boldsymbol{\theta}^\mathsf{tr}_u)=\mathbb{E}_\ell\left[y_\ell-Q_{\boldsymbol{\theta}^\mathsf{tr}_u}(\boldsymbol{S}_u^{(\ell)},\boldsymbol{a}_u^{(\ell)})\right],
    \label{eq:loss}
\end{equation}
where $	y_\ell = r^{(\ell)}_u + \gamma\max_{\boldsymbol{a}}\;Q_{\boldsymbol{\theta}^\mathsf{ta}_u}(\boldsymbol{S}^{(\ell+1)}_u,\boldsymbol{a})$ with $\gamma$ being the discount factor.
Here $Q_{\boldsymbol{\theta}}(\boldsymbol{S},\boldsymbol{a})$ represents the Q-value for a given pair of state $\boldsymbol{S}$ and action $\boldsymbol{a}$ computed via a DNN of weight vector $\boldsymbol{\theta}$.
The update is done using stochastic gradient descent (SGD).
Note that this step is equivalent to the training phase of supervised learning in the sense that each experience becomes an individual training datapoint and the label is replaced by the reward.
Here, the weights of $\boldsymbol{\theta}^\mathsf{tr}_u$ are periodically copied to target network $\boldsymbol{\theta}^\mathsf{ta}_u$, with the length of this period as a design parameter.

\textbf{Experience generation:} By the O-RAN capability, the value of $N_\mathsf{n}$ can vary and impact the rate of experiences being collected to each agent, i.e., the number of experiences collected for a given amount of time varies by $N_\mathsf{n}$.
If $N_\mathsf{n}$ is too large, a sufficient size of data required to perform effective training may not be collected within a desired time period.
To resolve the issue and utilize time more efficiently, we exploit the architecture of O-RAN and introduce an internal experience-generating loop inside the near-RT RIC.
This internal loop is executed $L$ times to take additional $L$ hypothetical actions during a single near-RT loop.
In particular, once a \textit{real} experience is obtained via the $\ell$-th near-RT loop, we generate $L$ extra \textit{virtual} experiences by taking a random action and evaluating the corresponding reward for each internal loop.
We define the reward by the $l$-th internal loop of the $\ell$-th near-RT PA as
\begin{equation}
    r^{(\ell,l)}_{\mathsf{int},u}(p)=\left(1-\kappa^{(\ell,l)}_u/\kappa_\text{max}\right)r^{(\ell)}_u(p),
    \label{eq:int_reward}
\end{equation}
where $\kappa^{(\ell,l)}_{u}=\left|\sum_{t=1}^{T_p}\left(\sum_{k\in\mathcal{K}^\textsf{DU}_u}(\boldsymbol{\phi}^{(N_\mathsf{n}\ell)}_{u,t})^\text{H}\boldsymbol{x}^{(\ell,l)}_k-\left\lfloor\frac{|\mathcal{K}^\mathsf{DU}_u|}{T_\text{p}}\right\rfloor\right)\right|$ is the penalty for having more than necessary number of users sharing the same pilot sequence and $\kappa_\text{max} = 2|\mathcal{K}^\mathsf{DU}_u|(T_p-1)/T_p$ is the maximum penalty obtainable.
Integrating this internal loop alongside near-RT PA, we can generate $L$ more experiences to accelerate the convergence of our scheme and train our DNNs to favor sequence combinations that have more evenly spread number of users across $T_p$ sequences.

\subsection{Iterative Codebook Search (CS) Algorithm}\label{ssec:methodology_codebook}

We describe our CS algorithm that is designed to work with the PA scheme in Sec.~\ref{ssec:methodology_framework}.
As each agent assigns pilots to its local users using the codebook $\mathcal{T}^{(i)}_u$, CS is iteratively conducted so that the final set of $U$ codebook sets, when combined with our PA solution, suppresses the PC to the minimum degree.
We detail each of the steps in the following.

First, we assign each agent $u\in\mathcal{U}$ with an identical codebook, i.e., $\mathcal{T}_1^{(0)}=\mathcal{T}_2^{(0)}=\cdots=\mathcal{T}_U^{(0)}$, and initiate our PA scheme without CS to ensure that the agents first learn and improve their PA only based on the interference resulted from pilot sharing.
We design our algorithm to begin its iterative CS only after the learning on PA is stabilized so that the PA and CS do not impair each other from converging.
We determine the PA of agent $u$ to be stable when the state $\boldsymbol{S}^{(\ell)}_u$ remains unchanged over $N_\mathsf{cs}$ near-RT loops.
Once the agent $u$ has given the same PA for $N_\mathsf{cs}$ consecutive times at the end of near-RT loop $\ell^\star_u$, the agent is perceived as stable and becomes subject for CS.
Note that $\ell^\star_u$ is likely to vary for each agent due to our decentralized PA framework.

If we design our agents to conduct CS in parallel, it becomes difficult to accurately evaluate a codebook as multiple actions simultaneously affect the environment.
Hence, we propose to have each agent take a turn and conduct CS while the rest of agents is paused from the search.
To implement a such design, we define an operation called the CS run in which an isolated CS is conducted for each agent $u \in \mathcal{U}^{(v)}_\mathsf{cs}$, where $\mathcal{U}^{(v)}_\mathsf{cs}$ is the set of agents subject for CS during the $v$-th CS run.
For each isolated search, the following steps are performed.

Suppose it is the turn of the $w$-th element of $\mathcal{U}^{(v)}_\mathsf{cs}$, denoted by $u_{v,w}$, to perform the isolated CS, where $w=1,2,\ldots,|\mathcal{U}^{(v)}_\mathsf{cs}|$.
We first define $\ell_{v,w}$ to be the near-RT loop in which the agent $u_{v,w}$ begins its search.
We also let $N_\mathsf{s}$ define the number of near-RT loops to be spent for codebook evaluation.
During the first $N_\mathsf{s}$ near-RT loops (i.e., $\ell_{v,w} \leq \ell < \ell_{v,w} + N_\mathsf{s}$), the quality of current codebook matrix $\mathbf{T}^{\mathsf{old}}_{v,w} =[\boldsymbol{\phi}_{u_{v,w},1}^{(N_\mathsf{n}\ell_{v,w})},\boldsymbol{\phi}_{u_{v,w},2}^{(N_\mathsf{n}\ell_{v,w})},\ldots,\boldsymbol{\phi}_{u_{v,w},T_p}^{(N_\mathsf{n}\ell_{v,w})}]$ is evaluated by computing
\begin{equation}
    \bar{r}^{\mathsf{old}}_{v,w} = \frac{1}{N_\mathsf{s}}\sum^{N_\mathsf{s}-1}_{n=0}r_{u_{v,w}}^{(\ell_{v,w}+n)}(p),
    \label{eq:CS_reward_old}
\end{equation}
which is the average of the most $N_\mathsf{s}$ recent rewards collected at agent $u_{v,w}$ via our PA algorithm.
Note that~\eqref{eq:CS_reward_old} represents the quality of PA performed using the codebook $\mathcal{T}^{(N_\mathsf{n}\ell_{v,w})}_{u_{v,w}}$.

After obtaining~\eqref{eq:CS_reward_old}, the agent generates a $T_p \times T_p$ column-normalized zero-mean Gaussian random perturbation matrix $\mathbf{P}_{v,w}$ and computes the rotation matrix as $\mathbf{R}_{v,w} = \sqrt{1-\eta_{u_{v,w}}^2}\mathbf{I}_{T_p}+\eta_{u_{v,w}}\mathbf{P}_{v,w}$, where $\eta_{u_{v,w}} = 1-\frac{\ell_{v,w}-\ell^\star_{u_{v,w}}}{\nicefrac{N}{{N_\mathsf{n}}}-\ell^\star_{u_{v,w}}}$ is the perturbation degree designed to decrease with $\ell_{v,w}$ to obtain a converged solution.
Note that larger $\eta_{u_{v,w}}$ results in $\mathbf{R}_{v,w}$ with greater perturbation.

After acquiring $\mathbf{R}_{v,w}$, the agent rotates the current codebook to obtain a new codebook matrix
\begin{equation}
    \mathbf{T}^\mathsf{new}_{v,w} = \textit{proj}(\mathbf{R}_{v,w}\mathbf{T}^{\mathsf{old}}_{v,w}),
    \label{eq:CS_codebook_new}
\end{equation}
where $\textit{proj}(\cdot)$ is the projection function for which we use the Gram-Schmidt orthogonalization algorithm~\cite{Golub96}.
The set of $T_p$ columns in $\mathbf{T}^\mathsf{new}_{v,w}$ is then used as a new codebook for agent $u_{v,w}$ during the next $N_\mathsf{s}$ near-RT loops (i.e., $\ell_{v,w} + N_\mathsf{s} \leq \ell < \ell_{v,w} + 2N_\mathsf{s}$).
After these $N_\mathsf{s}$ near-RT loops, where a set of $N_\mathsf{s}$ rewards using the new codebook are collected by our PA algorithm, the agent computes
\begin{equation}
    \bar{r}^{\mathsf{new}}_{v,w} = \frac{1}{N_\mathsf{s}}\sum^{2N_\mathsf{s}-1}_{n=N_\mathsf{s}}r_{u_{v,w}}^{(\ell_{v,w}+n)}(p),
    \label{eq:CS_reward_new}
\end{equation}
to evaluate the quality of the new codebook.
At this point, agent $u_{v,w}$ has evaluated~\eqref{eq:CS_reward_old} and~\eqref{eq:CS_reward_new} from using two different codebooks $\mathbf{T}^\mathsf{old}_{v,w}$ and $\mathbf{T}^\mathsf{new}_{v,w}$, respectively, and determines which codebook to keep by the end of search using the following criterion
\begin{equation}
\hspace{-10mm}\mathbf{T}_{u_{v,w}}^{(N_\mathsf{n}(\ell_{v,w}+2N_\mathsf{s}))} =
\begin{cases}
    \mathbf{T}^\mathsf{new}_{v,w} & \text{if } \; \bar{r}^{\mathsf{new}}_{v,w}  > \bar{r}^{\mathsf{old}}_{v,w}, \\
    \mathbf{T}^\mathsf{old}_{v,w} & \text{otherwise}.
\end{cases}
\label{eq:CS_determine}
\end{equation}

{\footnotesize
\begin{algorithm}[h]
    \caption{Proposed Codebook Search (CS) Scheme}
    \label{alg:PA-CS}
    {\small
    \textbf{Input:} Pilot length $T_p$, number of consistent PAs required for stability $N_\mathsf{cs}$, codebook evaluation interval $N_\mathsf{s}$, number of RT loops $N$, set of agents $\mathcal{U}$ \\
    Initialize CS run index $v=0$, set of agents subject for CS $\mathcal{U}^{(v)}_\mathsf{cs}=\varnothing$, the counter for agent $u$ $a_u=0$, $CS_\mathsf{run} = 0$, and $CS_\mathsf{iso} = 0$; assign identical codebook for all $u \in \mathcal{U}$; capture $\boldsymbol{S}^{(0)}_u$ using~\eqref{eq:state_entry} \\
    \For{$\ell=1$ to $N$}{
        \For{$u\in\mathcal{U}$}{
            Capture $\boldsymbol{S}^{(\ell)}_u$ using~\eqref{eq:state_entry} \\
            \textbf{if} $\boldsymbol{S}^{(\ell)}_u=\boldsymbol{S}^{(\ell-1)}_u$ \textbf{then} $a_u = a_u + 1$ \textbf{else} $a_u = 0$; \textbf{if} $a_u = N_\mathsf{cs}$ \textbf{then} $\ell^\star_{u} = \ell$ \\
        }
        \If{$CS_\mathsf{run} = 0$}{
            $\mathcal{U}^{(v)}_\mathsf{cs}=\{u\in\mathcal{U}|\ell^\star_{u}<\ell\}$; \textbf{if} $|\mathcal{U}^{(v)}_\mathsf{cs}|>0$ \textbf{then} $w=1$ and $CS_\mathsf{run} = 1$             
        }
        \If{$CS_\mathsf{run} = 1$}{
            \textbf{if} $CS_\mathsf{iso} = 0$ \textbf{then} $\ell_{v,w} = \ell$; $CS_\mathsf{iso} = 1$ \\
            \If{$CS_\mathsf{iso} = 1$}{
                \textbf{if} $\ell = \ell_{v,w} + N_\mathsf{s} - 1$ \textbf{then}
                    compute $\bar{r}^{\mathsf{old}}_{v,w}$ using~\eqref{eq:CS_reward_old}; apply new codebook $\mathbf{T}^\mathsf{new}_{v,w}$ using~\eqref{eq:CS_codebook_new} \\
                \If{$\ell = \ell_{v,w} + 2N_\mathsf{s} - 1$}{
                    Compute $\bar{r}^{\mathsf{new}}_{v,w}$ using~\eqref{eq:CS_reward_new}; decide codebook using~\eqref{eq:CS_determine}; $w = w + 1$ and $CS_\mathsf{iso} = 0$ \\
                }
            }
            \textbf{if} $w > |\mathcal{U}^{(v)}_\mathsf{cs}|$ \textbf{then} $v = v + 1$; $CS_\mathsf{run} = 0$ \\
        }
    }
    \textbf{Output:} Rotated codebook $\mathcal{T}^{(N)}_u,\forall u\in\mathcal{U}$
    }
\end{algorithm}
}

As the CS described above runs for each agent in $\mathcal{U}^{(v)}_\mathsf{cs}$, total $2N_\mathsf{s}|\mathcal{U}^{(v)}_\mathsf{cs}|$ near-RT loops are spent to complete the CS run $v$.
For every run, each agent tries a new codebook generated using a random rotation and decides to keep whichever codebook that yields higher reward.
The algorithm starts its very first CS run at $\ell=\min_{u\in\mathcal{U}} \ell^\star_u$ and continuously conducts each subsequent CS run.
By changing the codebook only when it is determined to be better, the algorithm proceeds to find the best set of $U$ codebooks that minimizes the degree of PC.
Note that, in order to evaluate the codebooks, our CS scheme utilizes the reward $r^{(\ell)}_u(p)$, which is obtained during our PA scheme.
Therefore, no additional information needs to be collected  the O-DUs to conduct the CS.
The overall procedure for our CS scheme is summarized in Alg~\ref{alg:PA-CS}.

\section{Numerical Evaluation}\label{sec:simulation}

In this section, we evaluate our pilot assignment (PA) scheme under O-RAN CFmMIMO channel estimation scenarios with various system parameters.
We analyze both channel estimation performance and computational complexity to discuss the scalability and practicality of our method.
In addition, we compare the performance of our proposed approach against different baselines which include~\cite{Liu20_1,Buzzi21} among others.

\subsection{Simulation Setup, Performance Metrics, and Baselines}\label{ssec:simulation_setup}

\begin{figure}[t]
    \centering
    \includegraphics[width=0.6\linewidth]{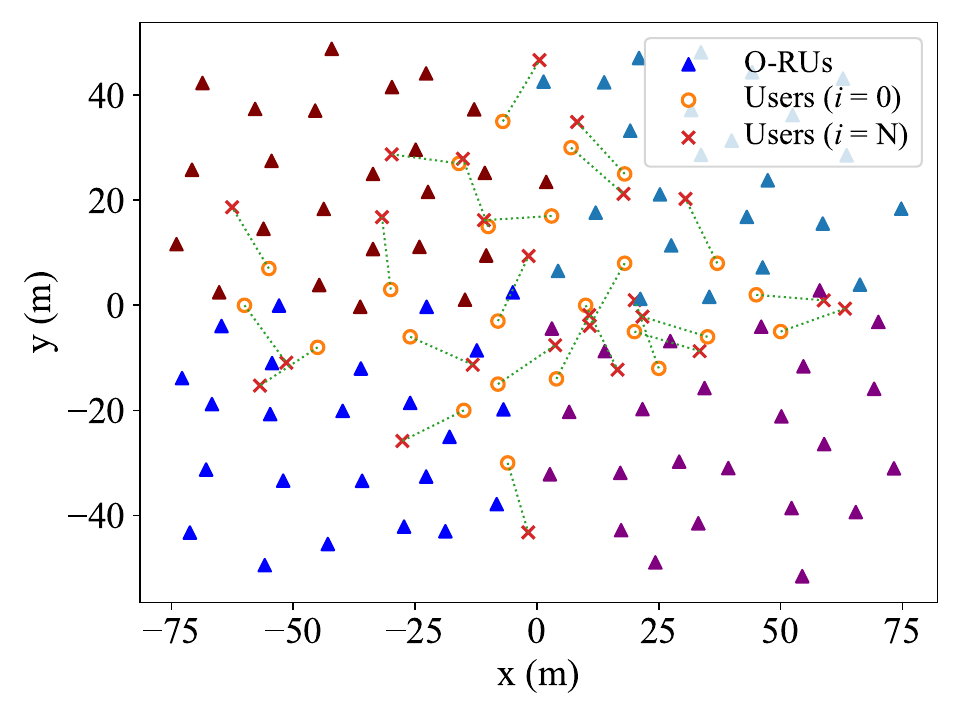}
    \caption{Geographical layout of O-RAN CFmMIMO with $U=4$, $M=96$, and $K=24$. O-RUs connected to the same O-DU have the same color. Each user moves from the initial (circle) to the final position (cross) in 10 seconds.}
    \label{fig:sim_cellfree_layout}
\end{figure}

We consider different combinations of O-DUs $(U = 4)$, single-antenna O-RUs $(M = 96)$, and single-antenna users $(K \in \{24,36\})$ placed in an area of $100\text{ m} \times 150\text{ m}$ geometry to create O-RAN CFmMIMO systems.
We assume the same number of O-RUs connected to each O-DU (i.e., $|\mathcal{M}^\mathsf{DU}_u|=\frac{M}{U},\forall u$) and the same number of users paired with each agent in the near-RT RIC (i.e., $|\mathcal{K}^\mathsf{DU}_u|=\frac{K}{U},\forall u$).
We set a  channel estimation interval $T_e=1$ ms, implying our O-RAN RT loop occurs once every $1$ ms.
Each scenario is simulated with a maximum $N=10000$ RT loops, which corresponds to $10$ seconds with $T_e=1$ ms.
We assume $N_\mathsf{n}=10$ RT loops occur per O-RAN near-RT loop and $L=9$ internal experience generation per near-RT loop unless stated otherwise.
For mobile scenarios, we generate initial ($i=0$) and final ($i=N$) positions for each user such that the velocity $v_k$ ranges from $0$ m/s (or $0$ km/h) to $1.4$ m/s (or $5$ km/h).
Then, for each $i=0,1,\ldots,N$, the position of each user is updated according to $v_k$. 
Such a mobile scenario for $96 \times 24$ CFmMIMO (where $M\times K$ refers to $M$ O-RUs and $K$ users) with $U=4$ O-DUs (equivalently, $U=4$ agents in the near-RT RIC) is demonstrated in Fig.~\ref{fig:sim_cellfree_layout}.
The large-scale fading factor $\beta^{(i)}_{km}$, $\forall k,m$ is assumed to follow the 3GPP urban-micro line-of-sight pathloss model~\cite{3GPP} with carrier frequency $f_c=2$ GHz, O-RU height of $10$ m, and user height of $1.5$ m.
We consider a pilot length of $T_p=4$ and a RU cluster size of $M^\mathsf{UE}_k=8,\forall k$ unless stated otherwise.
For our codebook search (CS) scheme, we consider an agent to be stable if the PA is consistent for $N_\mathsf{cs}=100$ consecutive times and assume the codebook evaluation interval $N_\mathsf{s}=5$.

We use the same DQN design for all agents: one convolutional neural network (CNN) with $32$ kernels of size $|\mathcal{K}^\mathsf{DU}_u|\times T_p$ followed by two fully connected layers of width $|\mathcal{K}^\mathsf{DU}_u|T_p$.
All layers use ReLU activation and the Adam optimizer with learning rate of $0.001$.
The discount factor for the weight update is set $\gamma=0.5$.
We also set the size of replay memory $D_\mathsf{m}=1000$ and train the neural network using $D_\mathsf{b}=128$ samples per minibatch.
The train network weights are updated via SGD and synchronized with the target network whenever $200$ and $400$ new additional experiences are stored in the replay memory, respectively.
We implement $\epsilon$-greedy action-selection~\cite{RL} with the probability of selecting a random action in the $\ell$-th near-RT loop computed as $\epsilon_\ell = e^{-(\Gamma/N)N_\mathsf{n}\ell}$, where $\Gamma=15$ is the scaling factor.

We now describe the baseline methods for performance comparison. 
We first consider a random assignment strategy (PA-RA) where pilots are assigned randomly for each user.
The strategy does not impose any complexity but yields mediocre channel estimation performance.
We also consider an exhaustive method (PA-ES) where the entire $T_p^K$ combinations of pilots are searched to find the PA having the lowest MSE, which is evaluated using $\beta_{km}$ and $\sigma^2$ assumed to be known a priori.
PA-ES provides the best MSE performance but is considered impractical in terms of computational complexity as the search space exponentially increases with the number of users.
We also consider two PA algorithms in the recent literature: PA strategies using Tabu-search~\cite{Liu20_1} and Hungarian~\cite{Buzzi21} methods.
To solve our PA problem, we design the algorithms to utilize sum-MSE as the metric.
The sum-MSE expression is a function of the assigned pilots and therefore provides an effective metric to optimize the PA.
Tabu-search-based PA (PA-TS) utilizes the Tabu-search framework to find the MSE-minimizing pilot combination while the PA using the Hungarian algorithm (PA-HG) iteratively solves a reward matrix to find the PA solution.
Both require prior knowledge of~$\beta_{km}$~and~$\sigma^2$ and have computational complexity that becomes prohibitive as the number of users increases.
Note that these methods do not consider practical framework (e.g., decentralized PA) but simply rely on centralized processing, which makes them hard to integrate into O-RAN architecture.
Also, they do not take the user mobility into account and fail to adapt to the change imposed by the time-varying dynamics.

We next discuss our PA scheme to be simulated for detailed evaluation.
We conduct the learning process described in Sec.~\ref{ssec:methodology_framework} with inter-DU message passing (PA-DRL+MSG), i.e., $\widetilde{p}_u^{(\ell)}$ is computed by each O-DU and transferred to the agent.
In addition, we apply the CS scheme described in Sec.~\ref{ssec:methodology_codebook} along with PA-DRL+MSG (PA-DRL+MSG+CBS) to assess the improvement brought by adjusting the codebook orientation across O-DUs.
As our PA scheme is specifically tailored to the O-RAN architecture, practical implementation with scalable computation is possible.
Since we base our learning on the DRL framework, which offers training that is adaptive to the dynamic environment, and conduct CS that checks the real-time observation, our PA scheme can reflect the user mobility.

We evaluate the performance of our proposed PA scheme over two different metrics: (i) the sum-MSE defined for the objective function in~$\boldsymbol{\mathcal{P}}_u$, i.e., $\sum_{k\in\mathcal{K}}\mathsf{MSE}^{(i)}_k$, and (ii) the runtime it takes to obtain the converged MSE.
For the numerical results, we run each scenario $50$ times and take their average to make our analysis statistically significant.
In each run, we use the same O-RU topology but randomize the locations of $K$ users.
All the algorithms were implemented in Python and tested on hardware with a Tesla T4 GPU and 12.7 GB RAM.

\subsection{Performance of O-RAN CFmMIMO}\label{ssec:simulation_ORAN}

\subsubsection{Impact of PA on channel estimation}
\begin{figure}[t]
    \centering
    \begin{subfigure}[!h]{0.49\textwidth}
        \centering
        \includegraphics[width=.85\linewidth]{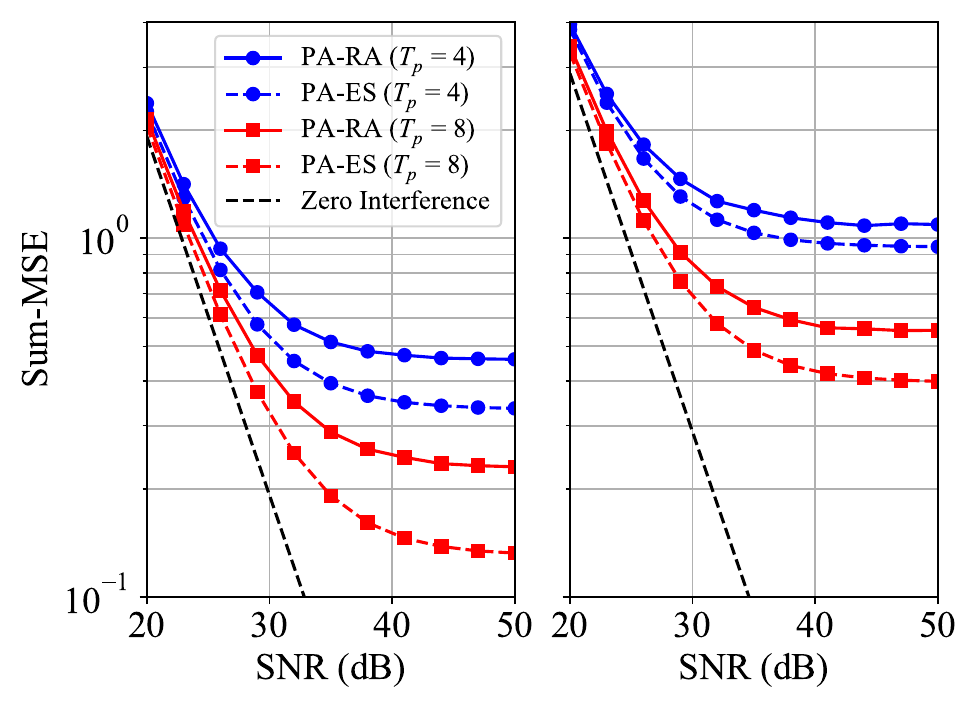}
        \vspace{-3mm}
        \caption{Sum-MSE vs. SNR with $K=24$ (left) and $K=36$ (right).}
        \label{fig:system_analysis}
    \end{subfigure}
    \begin{subfigure}[!h]{0.49\textwidth}
        \centering
        \includegraphics[width=.85\linewidth]{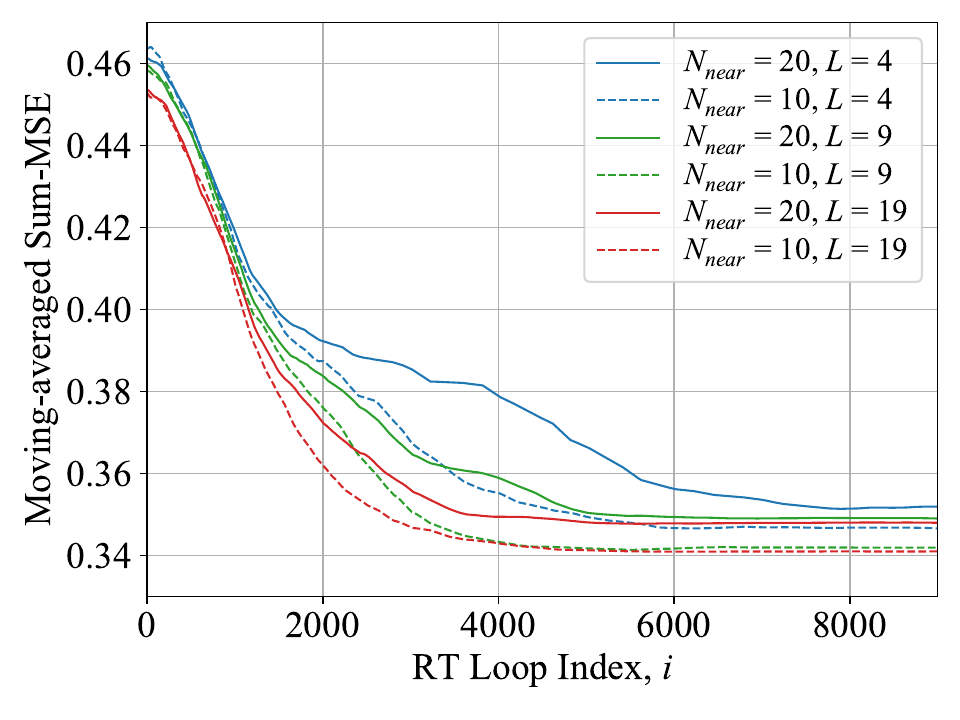}
        \vspace{-3mm}
        \caption{Sum-MSE vs. RT loop with $K=24$.}
        \label{fig:DRL_analysis}
    \end{subfigure}
    \vspace{0.5mm}
    \caption{Sum-MSE vs. SNR plot in terms of $T_p$ and $K$ (left) and sum-MSE vs. RT loop plot in terms of $N_\mathsf{n}$ and $L$ (right).}
\end{figure}

We first demonstrate the impact of PA on channel estimation in our O-RAN CFmMIMO system.
We provide sum-MSE versus signal to noise ratio (SNR) plots for different values of $T_p$ and $K$ in Fig.~\ref{fig:system_analysis} where we define SNR as $\frac{1}{\sigma^2}$.

Now we discuss several facts which are observed from the plots in Fig.~\ref{fig:system_analysis}.
First, we see that $T_p = 8$ yields lower MSE than $T_p = 4$. It is expected since the number of users sharing the same pilot tends to be smaller for larger $T_p$.
Next, for lower SNRs, the MSE gap between PA-RA and PA-ES is not significant since the noise dominantly contributes to channel estimation error.
However, as SNR increases, interference due to PC becomes more dominant and forces an error floor, making the curves almost horizontal.
For the case of $50$ dB SNR, we find that with $T_p=4$ and $K=24$, optimizing PA can reduce the sum-MSE up to 27\%.
For the remaining experiments, we use SNR of $50$ dB to focus on the interference-limited regime.

\subsubsection{Impact of O-RAN parameters} We assess the impact of O-RAN-dependent system parameters on the performance of our PA scheme.
The sum-MSE performance curves (moving-averaged with a window size of $500$) of PA-DRL+MSG over the O-RAN RT loop for different values of $N_\mathsf{n}$ and $L$ are shown in Fig.~\ref{fig:DRL_analysis}.
Recall that $N_\mathsf{n}$ is the number of RT loops for a single near-RT loop, and $L$ is the number of extra experiences generated per near-RT loop by the agent.
Both $N_\mathsf{n}$ and $L$ are dependent on the capability of O-RAN in which CFmMIMO network is built.

Now, we make the following observations from Fig.~\ref{fig:DRL_analysis}.
First, regardless of the parameter values, our scheme shows stabilized (i.e., converged) sum-MSE performance, which verifies the effectiveness of our learning when implemented under O-RAN architecture. 
Second, a lower $N_\mathsf{n}$ yields improved MSE regardless of $L$.
Here, lower $N_\mathsf{n}$ implies more near-RT loops during the given number of RT loops, allowing agents to interact with the environment more frequently and take more actions to find better solutions.
Third, a higher $L$ (more internal loops) allows us to achieve greater sum-MSE reduction in earlier RT loops, validating that more experiences collected in replay memory within the same period are beneficial.
Thus, as the size of the dataset increases, our scheme is expected to find the PA faster with low sum-MSE.

\subsection{Performance Comparison Against Different Baselines}\label{ssec:simulation_baseline}

Now we assess our proposed PA scheme and compare its performance with several baselines over two metrics: channel estimation MSE and algorithm runtime.

\begin{figure}[t]
    \centering
    \begin{subfigure}[!h]{0.49\textwidth}
        \centering
        \includegraphics[width=.95\linewidth]{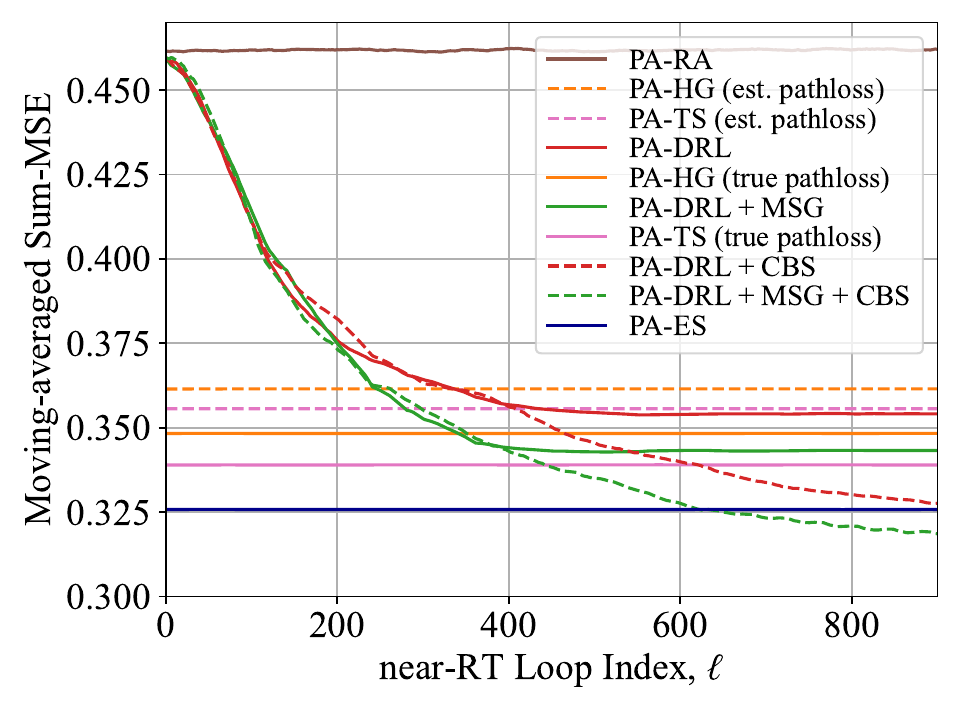}
        \vspace{-3mm}
        \caption{$K=24$ users}
        \label{fig:stationary_24}
    \end{subfigure}
    \begin{subfigure}[!h]{0.49\textwidth}
        \centering
        \includegraphics[width=.95\linewidth]{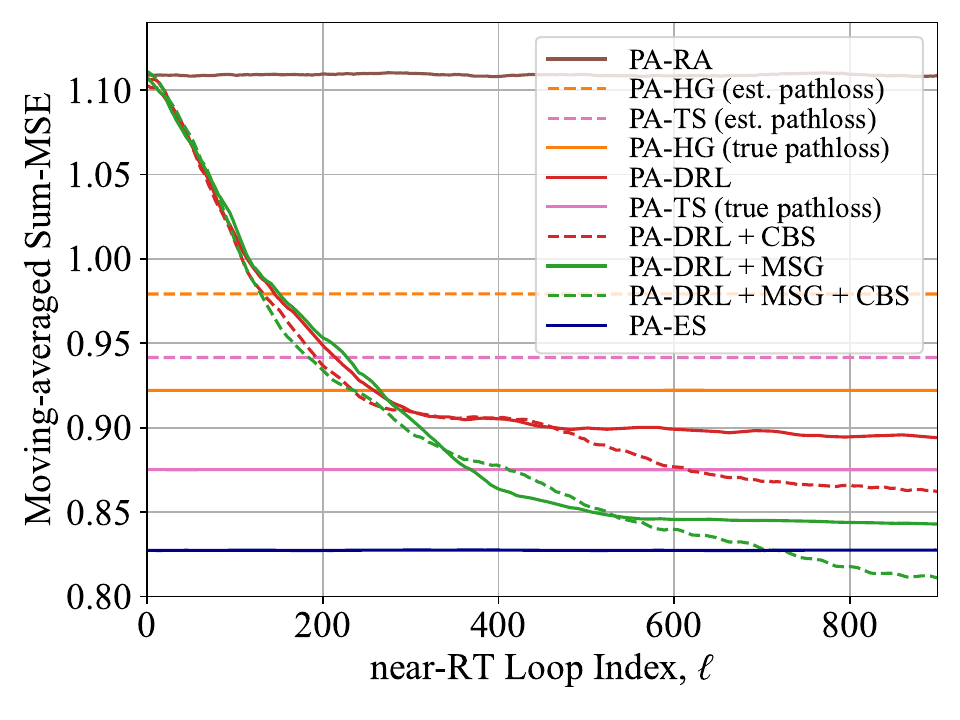}
        \vspace{-3mm}
        \caption{$K=36$ users}
        \label{fig:stationary_36}
    \end{subfigure}
    \caption{Sum-MSE performance of different PA schemes over 24 stationary users (left) and 36 stationary users (right).}
    \label{fig:stationary}
\end{figure}

\subsubsection{Comparison in MSE}
First, we consider static scenarios, i.e., $v_k=0, \forall k$.
The plots showing sum-MSE performance (moving-averaged with a window size of $500$) over RT loops for $K=24$ and $K=36$ are presented in Fig.~\ref{fig:stationary_24} and Fig.~\ref{fig:stationary_36}, respectively.
Note that the PA solutions obtained by PA-HG, PA-TS, and PA-ES required true pathloss information and were fixed for the entire RT loops.
Among these approaches, it is verified from both figures that PA-ES yields much better MSE performance than PA-TS and PA-HG.
We also considered the case where PA-HG and PA-TS are conducted using estimated pathloss, which yields a considerable performance gap compared to the case of using true pathloss knowledge.
The estimated pathloss is computed by averaging ten instantaneous power measurements from isolated signal transmissions, which yields around a $25\%$ error magnitude compared to true pathloss.
Given that these baselines require prior knowledge (preferably accurate) to achieve the given performance, our learning-based PA scheme, which does not impose such requirement, is still able to show competitive performance against them.
PA-DRL+MSG clearly outperforms PA-HG and PA-TS with estimated pathloss and provides comparable performance with the ones with true pathloss.
Once we utilize CS scheme, our proposed PA-DRL+MSG+CBS shows significant improvement and achieves better performance than PA-ES as a result of jointly optimizing both PA and codebook orientation.
In Table~\ref{tb:MSE_largeK}, we extend our sum-MSE evaluation up to $K=72$.
We observe that, regardless of $K$, the relative performance among the algorithms is preserved, which verifies that our proposed scheme obtains consistent improvements as the system size increases.

Note that we compare our proposed scheme only to centralized baselines for two reasons.
First, we aim to minimize the performance loss due to decentralization, and thus we can directly evaluate how well our algorithm performs in terms of MSE compared to centralized PA.
Second, to the best of our knowledge, there is no existing work on decentralized PA to have a valid comparison.

\begin{table}[!t]
\captionsetup{justification=centering, labelsep=newline}
\caption{Sum-MSE performance of various PA algorithms over different values of $K$}
\label{tb:MSE_largeK}
\centering
\begin{tabular}{|c|c|c|c|c|c|} 
    \hline
    Algorithm & $K=24$ & $K=36$ & $K=48$ & $K=60$ & $K=72$ \tabularnewline
    \hline
    PA-HG (true pathloss) & 0.348 & 0.949 & 1.705 & 2.765 & 4.179 \tabularnewline
    \hline
    PA-TS (true pathloss) & 0.339 & 0.875 & 1.657 & 2.677 & 4.049 \tabularnewline
    \hline
    PA-DRL + MSG + CBS  & \textbf{0.319} & \textbf{0.811} & \textbf{1.455} & \textbf{2.361} & \textbf{3.617} \tabularnewline
    \hline
\end{tabular}
\end{table}

\begin{figure}[t]
    \centering
    \begin{subfigure}[!h]{0.32\textwidth}
        \centering
        \includegraphics[width=1\linewidth]{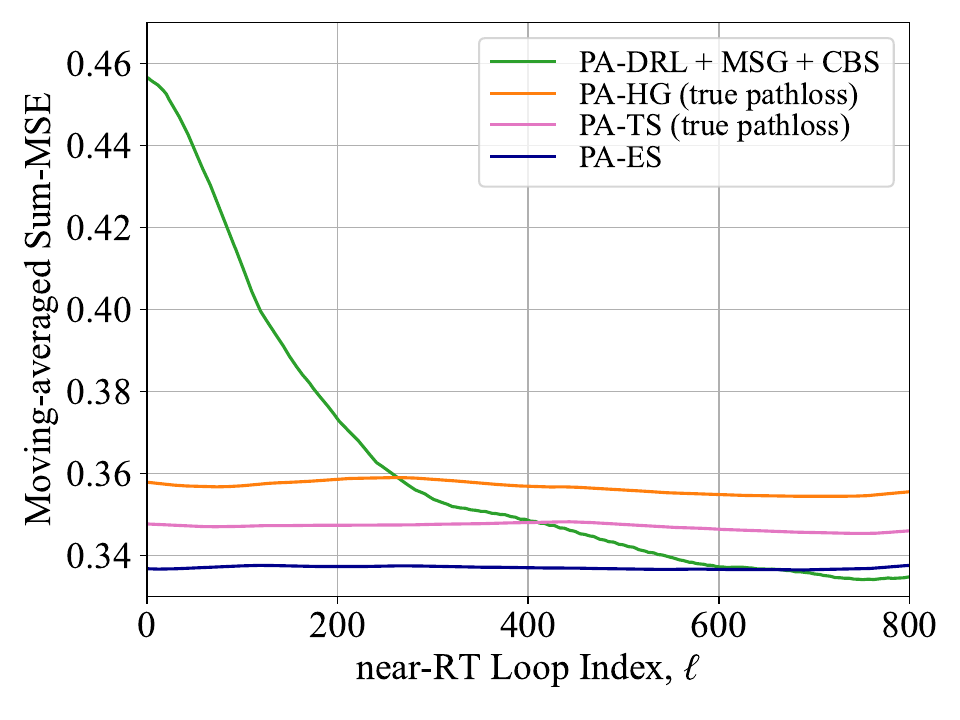}        
        \caption{Velocity = $1$ km/h}
        \label{fig:moving_24_1km}
    \end{subfigure}
    \begin{subfigure}[!h]{0.32\textwidth}
        \centering
        \includegraphics[width=1\linewidth]{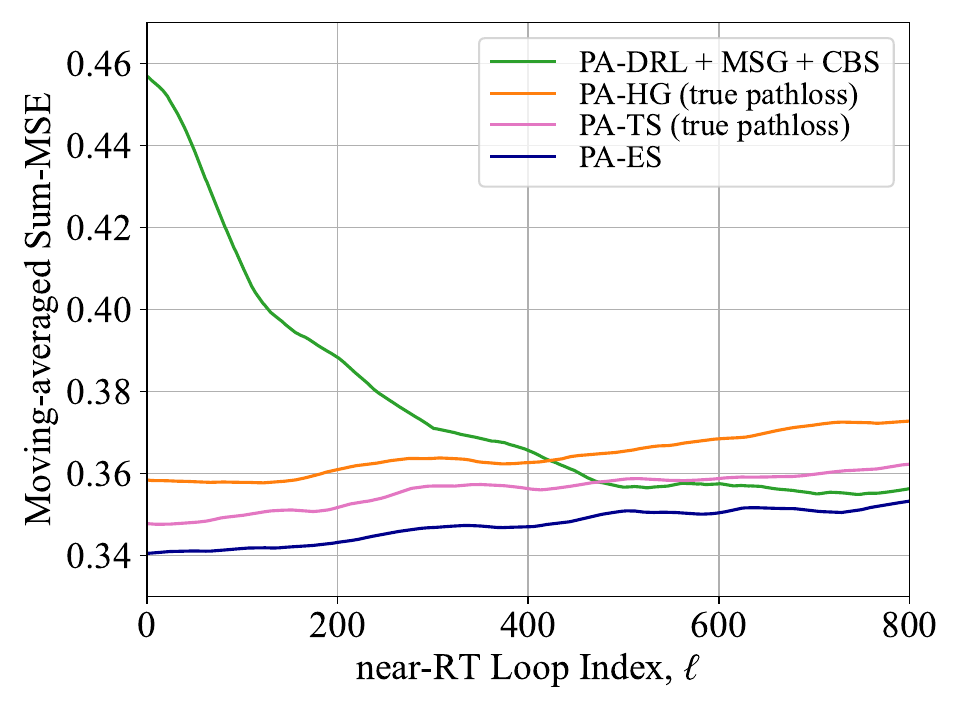}
        \caption{Velocity = $5$ km/h}
        \label{fig:moving_24_5km}
    \end{subfigure}
    \begin{subfigure}[!h]{0.32\textwidth}
         \centering
        \includegraphics[width=1\linewidth]{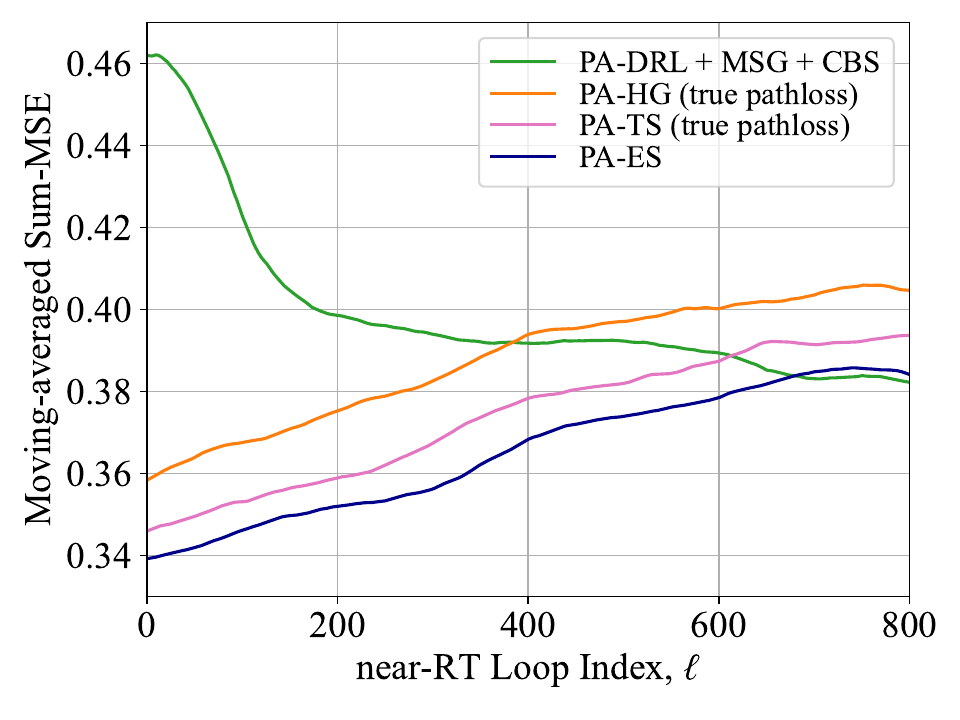}
        \caption{Velocity = $10$ km/h}
        \label{fig:moving_24_10km}
    \end{subfigure}
    \caption{MSE performance of different PA schemes over 24 mobile users with different velocities: $1$ km/h, $5$ km/h, and $10$ km/h.}
    \label{fig:moving_24}
\end{figure}

Next, we consider scenarios in which users move over time (i.e., $\beta^{(i)}_{km}$ changes over $i$, and $v_k > 0,\forall k\in\mathcal{K}$).
Fig.~\ref{fig:moving_24} shows the sum-MSE performance (moving-averaged with a window size of $500$) of different PA algorithms with $K=24$ evaluated at three different user velocities: $1$, $5$, and $10$ km/h.
The values of user velocity were selected so that the users still remain in the coverage area after their movement.
PA solution obtained by the baselines at the beginning (i.e, $i=0$) becomes less effective as time advances, showing a different degree of steady increase by the velocity.
Unlike the baselines, as our schemes make their decisions based on the real-time observations, in PA-DRL+MSG+CBS, PAs can be performed in an adaptive manner, maintaining its performance as shown in Fig.~\ref{fig:moving_24}.
Hence, our scheme can provide competitive performance with the prior knowledge-constrained baseline methods under a dynamic environment.

While our evaluation assumes ideal link connections, imperfect connection links are a significant factor for practical systems.
Hence, we extended our experiment to assume probabilistic link failures on O-FH and inter-DU connections and analyze their impact on the channel estimation performance. 
We observe that the overall sum-MSE increases with the increase of link failure probabilities.
Such a result is expected as failed connections prevent the agents from collecting necessary observations and computing accurate rewards.
For more details, see Appendix~\ref{app:A}.

To consider a wider range of scenarios, we also considered our experiment under three additional setups: non-uniform user distribution, Rician channel fading with different k-factor values~\cite{MIMO}, and correlated pathloss with different shadowing variance~\cite{Ngo17}.
We observe that the sum-MSE performance of our scheme and the baselines remains unchanged except for the case of increased shadowing.
With greater shadowing, the expected degree of PC increases, and this results in increased sum-MSE of channel estimation for all algorithms.
For more details, see Appendix~\ref{app:B}.

Overall, our scheme provides satisfactory performance in MSE as it exploits the decentralized architecture of O-RAN CFmMIMO via distributed learning and CS.

\subsubsection{Comparison in SE}
We evaluate uplink and downlink achievable SEs by computing  $\sum_{k\in\mathcal{K}}R^{\mathsf{u}}_{k}$ and $\sum_{k\in\mathcal{K}}R^{\mathsf{d}}_{k}$, respectively, for different PA algorithms.
The result is provided in Table~\ref{tb:sum_rate}, and we make the following observations.
\begin{table}[!t]
\captionsetup{justification=centering, labelsep=newline}
\caption{UL and DL SEs in bits/s/Hz for different PA algorithms.}
\label{tb:sum_rate}
\centering
\begin{tabular}{|c|c|c|c|c|}
    \hline
    \multirow{2}{*}{Algorithm} & \multicolumn{2}{c|}{$K=24$} & \multicolumn{2}{c|}{$K=36$} \\ \cline{2-5}
    & UL & DL & UL & DL \tabularnewline
    \hline
    PA-HG (true pathloss) & 9.79 & 9.03 & 7.65 & 7.10 \tabularnewline
    \hline
    PA-TS (true pathloss) & 9.89 & 9.11 & 7.81 & 7.23 \tabularnewline
    \hline
    PA-DRL + MSG + CBS & \textbf{10.18} & \textbf{9.40} & \textbf{8.06} & \textbf{7.55} \tabularnewline
    \hline
\end{tabular}
\end{table}
First, the SE performance with $K=36$ is lower than the one with $K=24$, which is resulted from increased user density imposing a larger degree of PC.
Second, the performance order we observe in Fig.~\ref{fig:stationary} is preserved for the sum-rate performance.
This verifies that improving channel estimation accuracy via effective PA results in the SE improvement in both uplink and downlink phases.

\subsubsection{Behavior comparison in relative runtime} 
\begin{figure}[!t]
    \centering
    \begin{subfigure}[!h]{0.32\textwidth}
        \centering
        \includegraphics[width=1\linewidth]{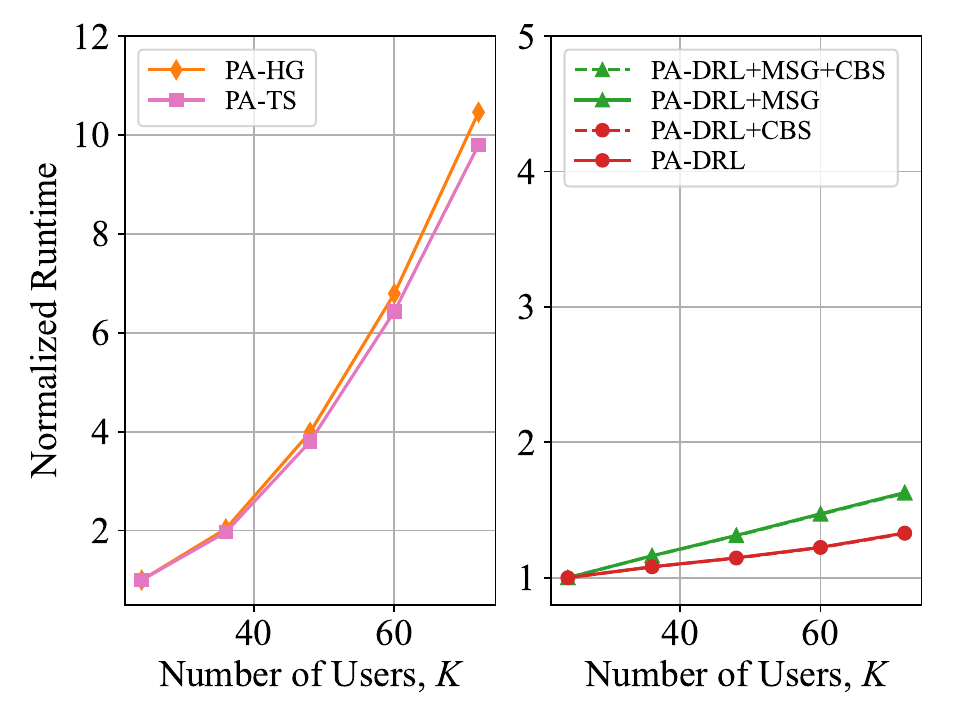}
        \caption{{Relative runtime for the baselines (left) and the proposed (right) over different $K$ values.}}
        \label{fig:complexity_user}
    \end{subfigure}
    \begin{subfigure}[!h]{0.32\textwidth}
        \centering
        \includegraphics[width=1\linewidth]{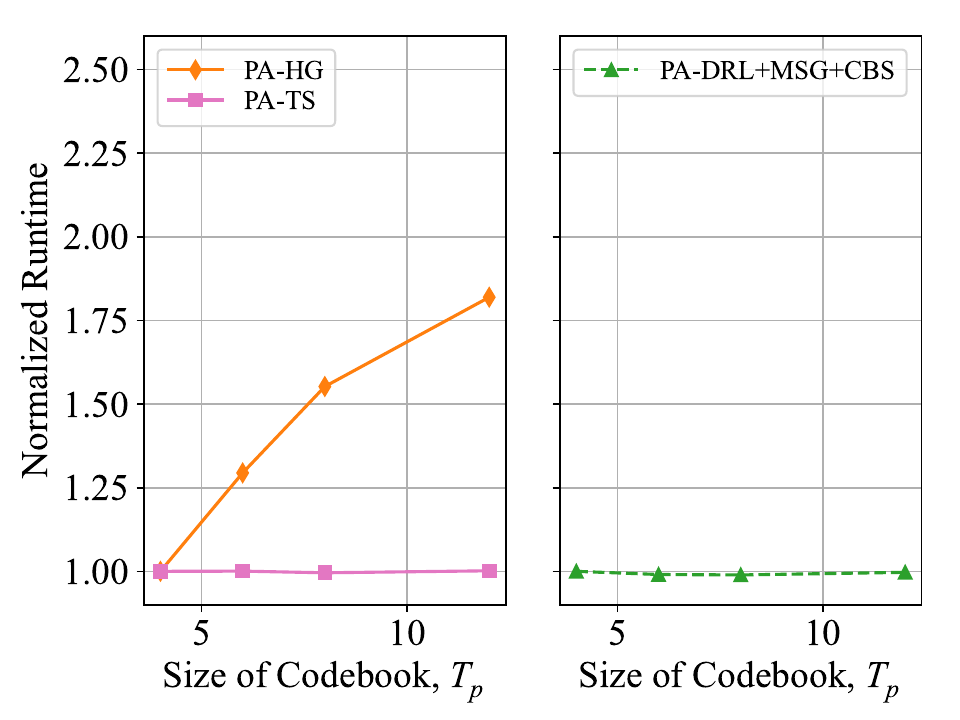}
        \caption{{Relative runtime for the baselines (left) and the proposed (right) over different $T_p$ values.}}
        \label{fig:complexity_codebook}
    \end{subfigure}
    \begin{subfigure}[!h]{0.32\textwidth}
        \centering
        \includegraphics[width=1\linewidth]{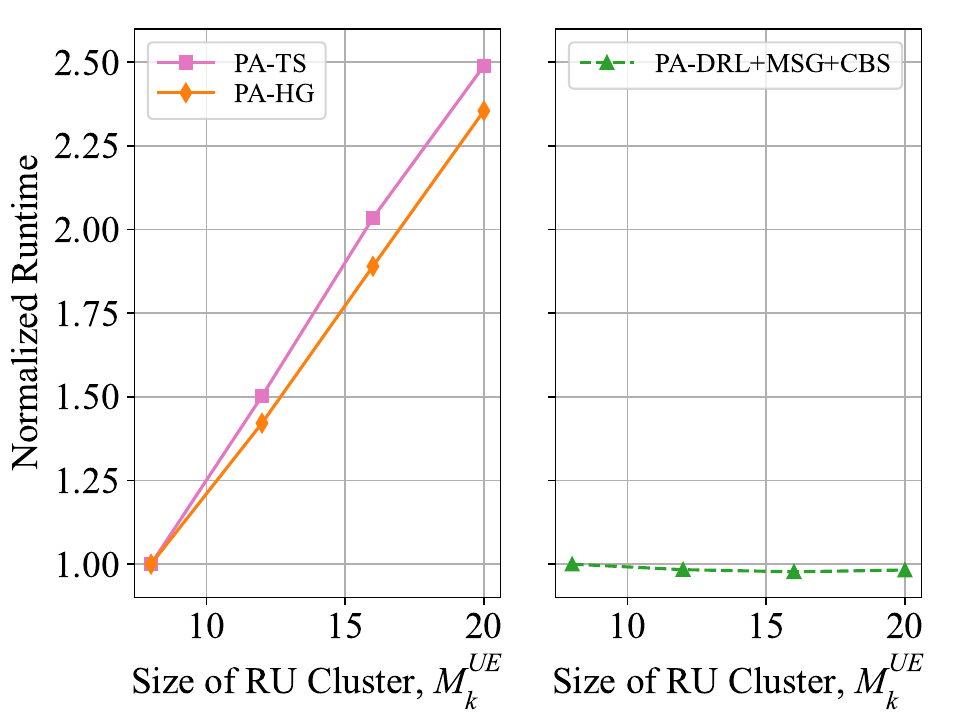}
        \caption{{Relative runtime for the baselines (left) and the proposed (right) over different $M^\mathsf{UE}_k$ values.}}
        \label{fig:complexity_cluster}
    \end{subfigure}
    \caption{Relative runtime measurements of various PA schemes over different system parameters. Measurements for each PA algorithm are normalized to the case of the smallest parameter value.}
    \label{fig:complexity}
\end{figure}

Now, we evaluate the computational complexity behavior of different PA algorithms.
In Fig.~\ref{fig:complexity_user}, we vary the number of users $K$ from $24$ to $72$ and measure the relative runtimes (i.e., the increase of runtime with respect to the case $K = 24$) of different PA methods.
We do not report absolute runtimes as our implementations do not factor in inter-node bandwidth, latency, and other implementation/hardware-specific factors that can impact head-to-head comparisons among the algorithms.
In Fig.~\ref{fig:complexity_user}, we see that both centralized PA-TS and PA-HG exhibit a polynomial increase in $K$.
Hence, these centralized algorithms can be rendered impractical when PA needs to be executed over a CFmMIMO network with a growing network size.
On the other hand, our PA algorithm shows a linear increase in the relative runtime.
Overall, combined with the MSE results, we see that our decentralized PA approach obtains advantages both in terms of pilot contamination and scalability.
The steady increase in runtimes from our PA scheme are due to the utilization of (i) O-RAN architecture where duration-varing tasks are distributed across the network and (ii) DNNs of fixed size which only perform a forward computation to determine each pilot update step over near-RT loop.
We observe a slight increase in runtime when we consider inter-DU messages into our PA scheme because generating a new set of messages imposes extra computations.
Note that our CS scheme barely adds any runtime as it utilizes the rewards already computed during our PA scheme.
We hence conclude that our low-complexity PA scheme is a scalable strategy that supports large-scale CFmMIMO systems.
As we have previously shown, our PA scheme provides consistently strong performance in terms of sum-MSE regardless of $K$, which highlights the scalability advantage of our approach, especially for large-scale systems.
Note that PA-ES, which is the best baseline in MSE minimization, requires an extreme amount of runtime as it searches over all $T_p^K$ combinations of PA.
On the other hand, PA-RA requires no extra runtime but shows much worse MSE performance than other PA schemes (Figs. \ref{fig:stationary_24} and \ref{fig:stationary_36}).

Next, we assess the runtime required to conduct PA algorithms over different values of $T_p$ (Fig.~\ref{fig:complexity_codebook}) and $M^\mathsf{UE}_k$ (Fig.~\ref{fig:complexity_cluster}), where we normalize the measurements in the same way as Fig.~\ref{fig:complexity_user}.
For varying $T_p$ (the size of codebook), only PA-HG shows undesirable behavior in complexity
since the size of the reward matrix used in the Hungarian algorithm depends on $T_p$.
With respect to $M^\mathsf{UE}_k$ (the size of RU cluster), both PA-TS and PA-HG display a linear increase.
Meanwhile, our proposed scheme provides consistent runtimes for both parameters, which verifies their scalability to support a network with large system parameters.

\section{Conclusion}\label{sec:conclusion}

In this paper, we developed a learning-based PA scheme for the decentralized CFmMIMO system framed in O-RAN.
We adopted O-RAN as a practical system architecture where distinct network functions and multi-timescale control loops efficiently govern the framework of our scheme.
After formulating the PA problem and designing the corresponding Markov game model, we developed a PA algorithm based on the MA-DRL framework.
We also developed a CS scheme that accelerates our learning-based PA in MSE-minimization without any significant additional complexities.
Compared to the state-of-the-art baselines, our approach provided satisfactory performance in terms of both channel estimation MSE and computational scalability.
Furthermore, unlike most of the existing PA strategies, our scheme does not require any prior channel knowledge. 

\appendices

\section{Analysis on Imperfect Connection Links}\label{app:A}

Connection failure is an important condition to consider for practical cell-free massive MIMO systems.
Hence, we reflect non-ideal O-FH and inter-DU connections into our pilot assignment framework as follows.

\textit{Inter-DU connection:} Inter-DU connection allows O-DUs to exchange information required to compute~\eqref{eq:pu_tilde}.
To model probabilistic connection failures, we introduce a binary variable denoted by $\varepsilon_{um}\in\{0,1\}$ to indicate the connection status between O-DU $u$ and the O-DU to which O-RU $m$ is connected.
We assume $\varepsilon_{um}\sim\text{Bernoulli}(1-P_\mathsf{d})$ with $0\leq P_\mathsf{d} \leq 1$ being the inter-DU connection failure probability.
Subsequently, to reflect the non-ideal connection, we modify~\eqref{eq:pu_tilde} as
\begin{align}
    \widetilde{p}^{(\ell)}_{u} &= \bar{p}^{(\ell)}_{u} + \underbrace{\sum_{k\in\mathcal{K}^\mathsf{DU}_u}\sum_{m\in\mathcal{M}^\mathsf{UE}_k\setminus\mathcal{M}^\mathsf{DU}_u} \varepsilon_{um}p^{(\ell)}_{km}}_{\text{via inter-DU connection}}.
    \label{eq:pu_tilde_modified}
\end{align}
Depending on the inter-DU connection status, some of the information required to reinforce the observation cannot be transferred.
Note that cases with $P_\mathsf{d}=1$ and $P_\mathsf{d}=0$ correspond to ``DRL" (with no inter-DU connection) and ``DRL+MSG" (with ideal inter-DU connection) in Section IV, respectively.

\textit{O-FH connection:} Each O-DU $u$ receives the observation made by O-RUs that are connected via an O-FH connection to compute $\bar{p}^{(\ell)}_{u}$ in~\eqref{eq:pu_tilde}.
In a similar way to the inter-DU connection failure model, we introduce a binary variable denoted by $\alpha_{m}\in\{0,1\}$ to indicate the connection status of O-RU $m$ to its O-DU.
We assume $\alpha_{m}\sim\text{Bernoulli}(1-P_\mathsf{f})$ with $0\leq P_\mathsf{f} \leq 1$ being the O-FH connection failure probability.
The expression of $\bar{p}^{(\ell)}_{u}$ is then modified as
\begin{equation}
    \bar{p}^{(\ell)}_{u} = \sum_{k\in\mathcal{K}^\mathsf{DU}_u}\sum_{m\in\mathcal{M}^\mathsf{UE}_k\cap\mathcal{M}^\mathsf{DU}_u} \alpha_{m}p^{(\ell)}_{km}.
\end{equation}
Depending on the O-FH connection status, some of the information required to completely evaluate $\bar{p}^{(\ell)}_{u}$ does not arrive at the O-DU.

\begin{figure*}[!t]
    \centering
    \minipage{0.49\textwidth}
        \includegraphics[width=1\linewidth]{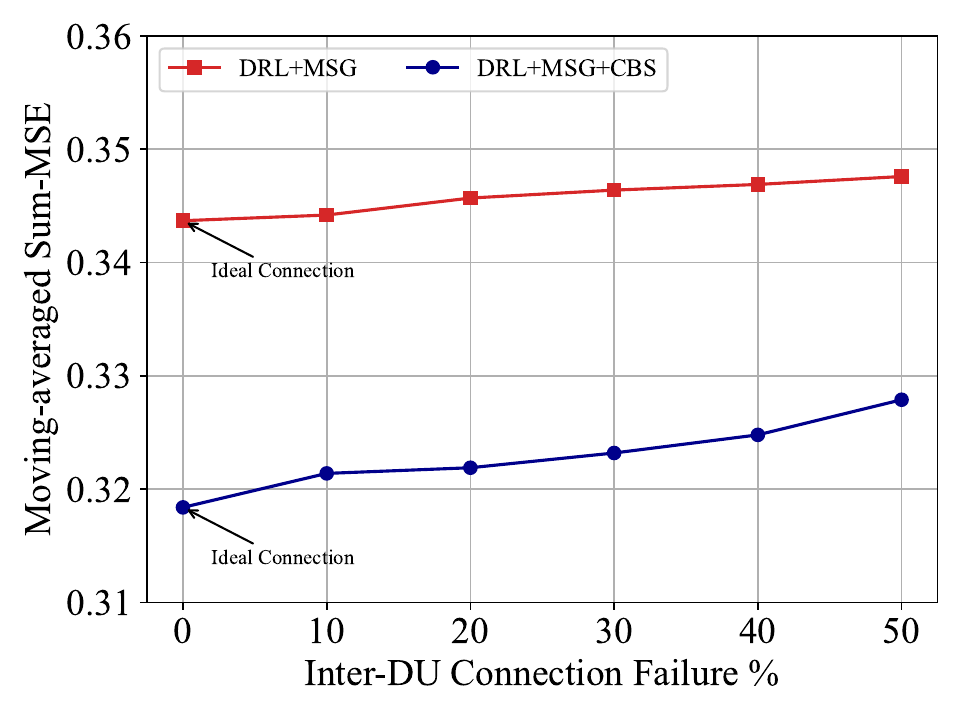}
        \caption{Sum-MSE vs. different inter-DU connection link failure probabilities $P_\mathsf{d}$. ($P_\mathsf{f}=0$)}
        \label{fig:DU_fail}
    \endminipage
    \hfill
    \minipage{0.49\textwidth}
        \includegraphics[width=1\linewidth]{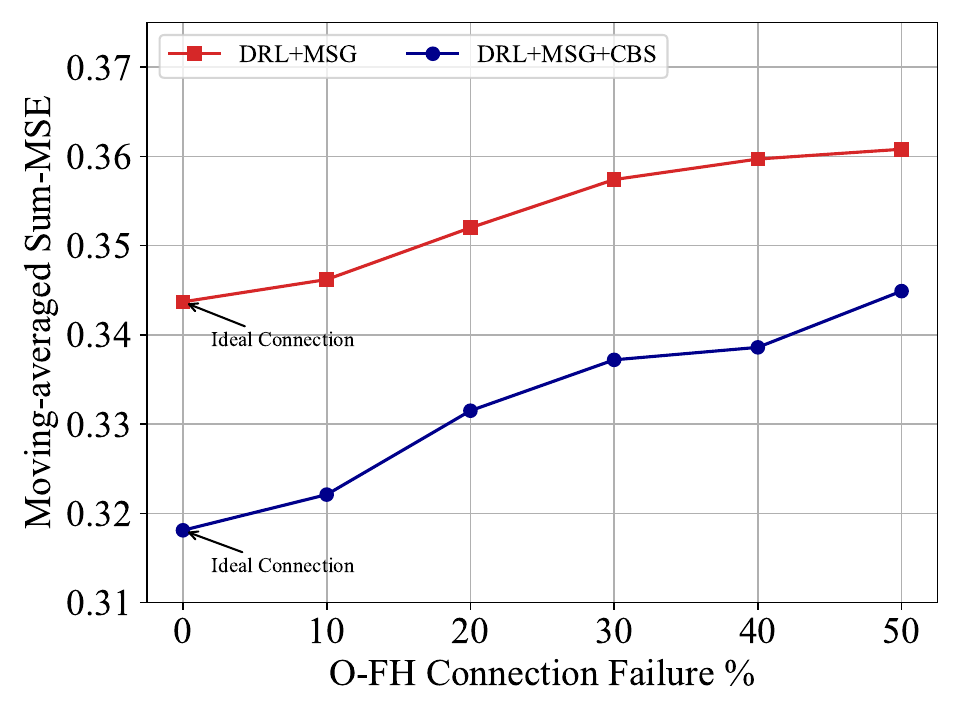}
        \caption{Sum-MSE vs. different O-FH connection link failure probabilities $P_\mathsf{f}$. ($P_\mathsf{d}=0$)}
        \label{fig:FH_fail}
    \endminipage
\end{figure*}

We now demonstrate the impact of having non-ideal connections on the O-FH and inter-DU connection.
We use the setting described in Section IV-A and make independent evaluations on $\alpha_m$ and $\varepsilon_{um}$ (i.e., we keep one variable fixed while varying the other variable).
Sum-MSE performance with $K=24$ obtained over different values of $P_\mathsf{d}$ and $P_\mathsf{f}$ are given in Figs.~\ref{fig:DU_fail} and~\ref{fig:FH_fail}, respectively.
We see that, for both inter-DU and O-FH connections, a higher failure probability results in a larger sum-MSE.
This is expected because imperfect connection links prevent the system from collecting information to accurately perceive the environment and degrade the efficiency of our solution.
Note that the performance of the ideal connection case matches the result presented in Fig.~\ref{fig:stationary}.

\section{Analysis with Additional Setups}\label{app:B}

\begin{figure*}[!h]
    \centering
    \minipage{0.49\textwidth}
        \includegraphics[width=1\linewidth]{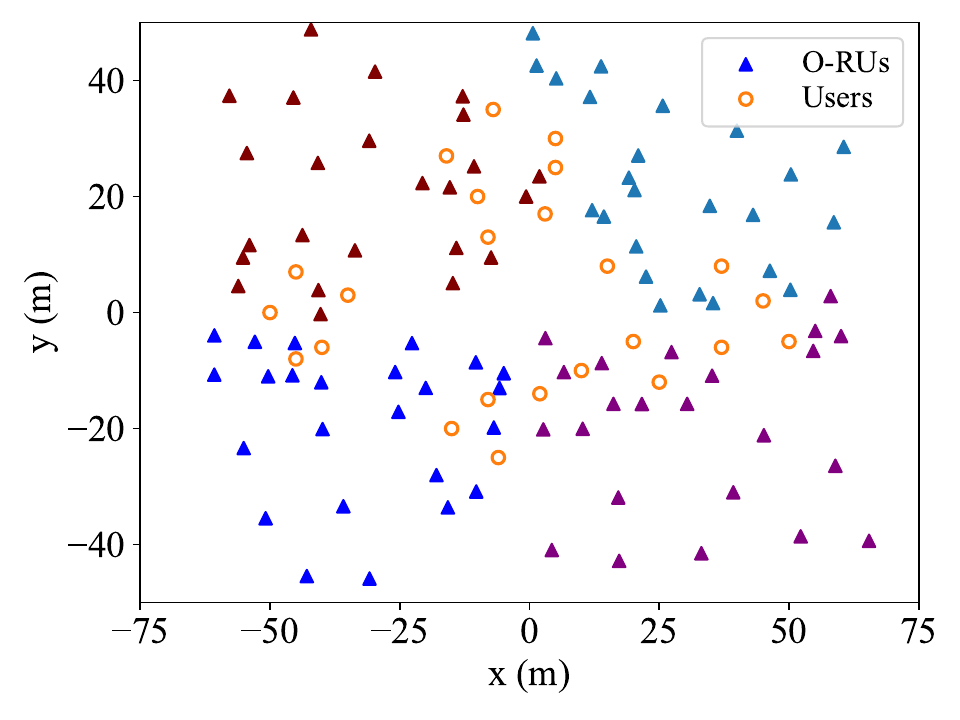}
        \caption{An example topology where O-RUs and users are placed via multiple high-density points.}
        \label{fig:dense_layout}
    \endminipage
    \hfill
    \minipage{0.49\textwidth}
        \includegraphics[width=1\linewidth]{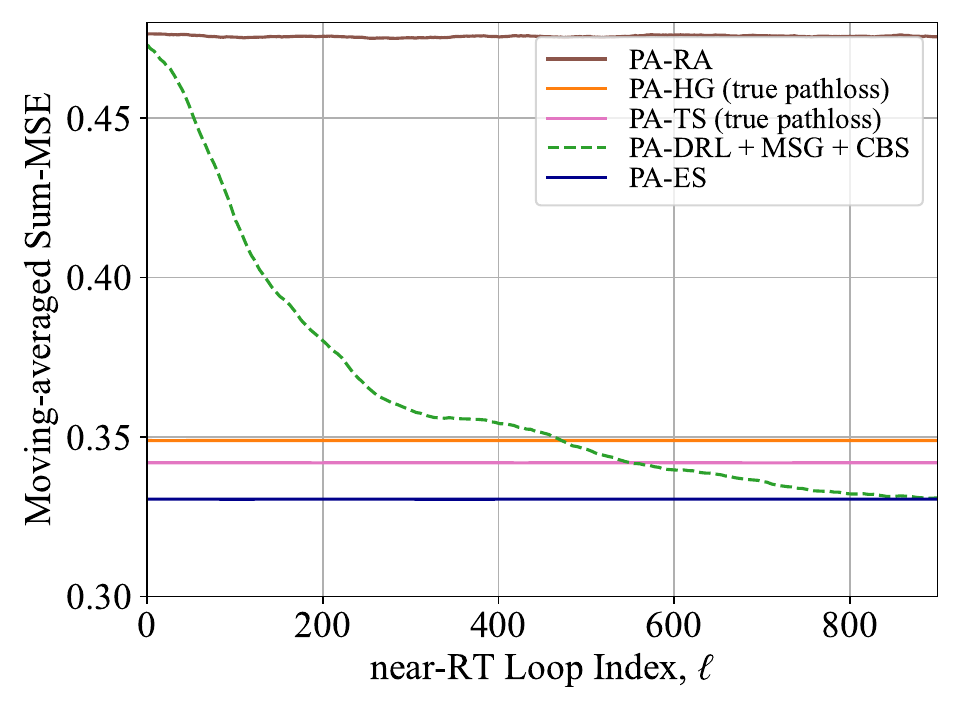}
        \caption{Sum-MSE performance of pilot assignment algorithms when a non-uniform distribution is used.}
        \label{fig:dense_MSE}
    \endminipage
\end{figure*}

Since \textit{user-centric connection}~\cite{Ammar22_1} is the key mechanism in a cell-free massive MIMO system,
following~\cite{Ngo17,Chen21_1,Attarifar18,Bjornson20_1}, we use uniform distribution to place O-RUs and users evenly in the area so that the users are likely to be surrounded by the O-RUs.
However, uniformly distributed layouts may not always be available in a practical scenario.
Hence, we generate scenarios where denser deployment of O-RUs and users is made on several points across the area and evaluate the sum-MSE performance.
An example layout and the corresponding sum-MSE plot are shown in Fig.~\ref{fig:dense_layout} and Fig.~\ref{fig:dense_MSE}, respectively.
We observe that the relative performance among the pilot assignment algorithms remains the same as compared to Fig.~\ref{fig:stationary}.

To consider various types of fading other than the one considered in Sec.~\ref{ssec:system_channel}, we experiment on the following channel conditions: (i) Rician fading~\cite{MIMO} through which both line-of-sight (LOS) and NLOS fading channels can be reflected and (ii) random correlated shadowing on the long-term pathloss~\cite{Ngo17}.
The Rican fading channel is generated based on the equation~\cite{MIMO}
\begin{equation}
    h_{km}^{(i)}=\sqrt{\frac{\kappa}{\kappa+1}}(1 \cdot e^{j\theta_{km}})+\sqrt{\frac{1}{\kappa+1}}\mathcal{CN}(0,1),
    \label{eq:Rician}
\end{equation}
where $\kappa$ and $\theta_{km}$ are the k-factor and uniform phase, respectively.
For the correlated shadowing, both the user and O-RU shadowing are assumed to follow a lognormal distribution with zero-mean and variance $\sigma^2_\mathsf{s}$.

\begin{figure*}[!h]
    \centering
    \minipage{0.49\textwidth}
        \includegraphics[width=1\linewidth]{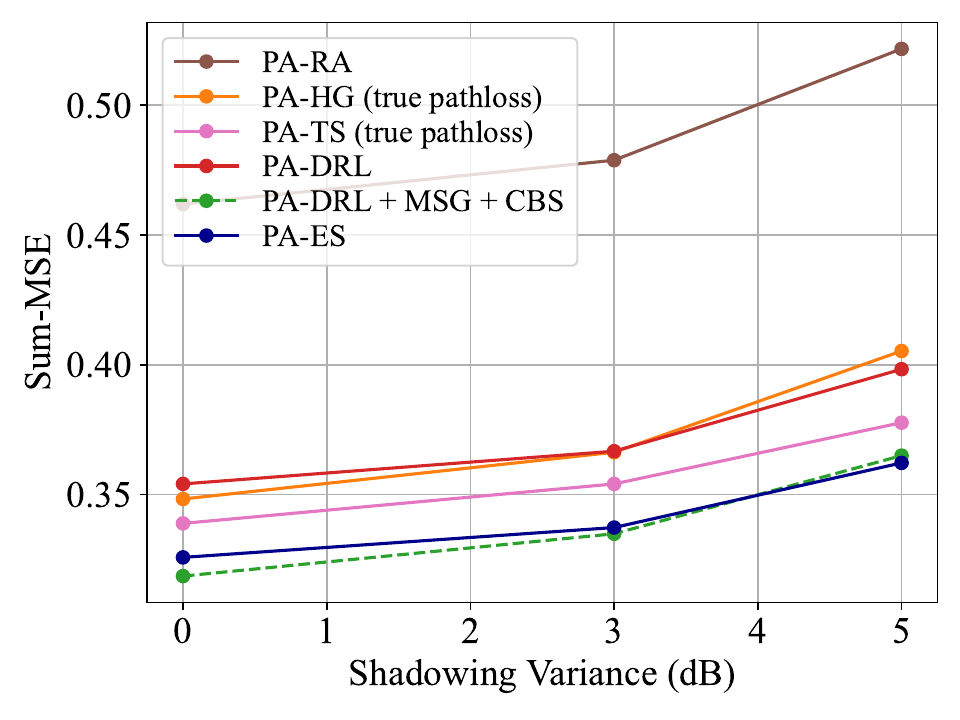}
        \caption{Sum-MSE performance of pilot assignment algorithms for different values of shadowing variance $\sigma^2_\mathsf{s}$.}
        \label{fig:shadowing}
    \endminipage
    \hfill
    \minipage{0.49\textwidth}
        \includegraphics[width=1\linewidth]{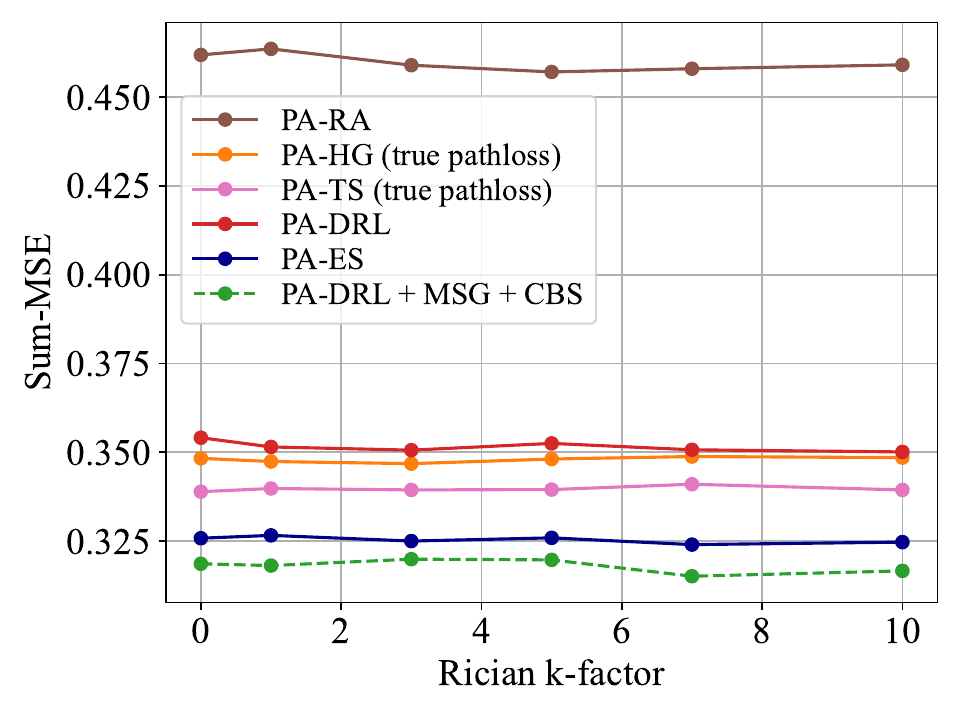}
        \caption{Sum-MSE performance of pilot assignment algorithms for different values of Rician k-factor $\kappa$.}
        \label{fig:rician}
    \endminipage
\end{figure*}

To numerically demonstrate the impact of having different channel conditions, we have conducted a set of simulations to evaluate the sum-MSE performance of pilot assignment algorithms under different values of $\sigma^2_\mathsf{s}$ and $\kappa$.
The simulation setup described in Section IV-A was used.
Figs.~\ref{fig:shadowing} and~\ref{fig:rician} show the change in sum-MSE performance for different values of shadowing variance and k-factor, respectively.

In Fig.~\ref{fig:shadowing}, we observe that the sum-MSE increases with the shadowing variance.
This is expected since imposing a multiplicative factor of greater variance to each pathloss increases their expected sum.
We also see from the figure that the relative performance among the pilot assignment algorithms does not change.
In other words, the presence of pathloss shadowing does not impact a certain algorithm in a specific way.
Meanwhile, the result in Fig.~\ref{fig:rician} shows that having our small-scale fading model follow a certain condition (i.e., LOS or NLOS) does not bring significant changes in the sum-MSE performance.
The considered algorithms (our learning-based approach as well as the baselines) aim to minimize the MSE defined as~\eqref{eq:MSE_4} of the revised manuscript;
since the expression is strictly dependent on the pathloss (i.e., the large-scale fading model) and noise variance terms, the performance of pilot assignment algorithms should remain stable against any variation in the small-scale fading model.

\bibliographystyle{IEEEtran}
\bibliography{IEEEfull,mybib}

\end{document}